\documentclass[aps,pra,twocolumn,superscriptaddress,amsmath]{revtex4-2}

\usepackage{bibunits}
\defaultbibliographystyle{long_bib} 

\defaultbibliography{main}

\usepackage{xcolor}
\usepackage{siunitx}
\usepackage{braket}
\usepackage{graphicx}% Include figure files
\usepackage{dcolumn}% Align table columns on decimal point
\usepackage{bm}% bold math
\usepackage{blindtext}
\usepackage{isotope}
\usepackage{hyperref}% add hypertext capabilities
\usepackage[normalem]{ulem}

\newcommand{\supplementarysection}{%
  \setcounter{figure}{0}% Reset figure counter
  \renewcommand{\thefigure}{S\arabic{figure}}% Prefix figure number with S
  \setcounter{section}{0}% Reset section counter
  \renewcommand{\thesection}{S\arabic{section}}% Prefix section number with S
  \setcounter{table}{0}% Reset section counter
  \renewcommand{\thetable}{S\arabic{table}}% Prefix section number with S
  \setcounter{equation}{0}% Reset equation counter
  \renewcommand{\theequation}{S\arabic{equation}}% Prefix equation number with S
  \section*{Supplementary Material}% Set supplementary section
  }
\begin{document}

\newcommand{\bohrradius}{\text{a}_0}
\newcommand{\potas}{\isotope[]{K} }
\newcommand{\potasBnospace}{$^{39}$K}
\newcommand{\potasB}{$^{39}$K }
\newcommand{\potasBB}{$^{41}$K }
\newcommand{\potasF}{$^{40}$K}
\newcommand{\cs}{\isotope[133]{Cs} }
\newcommand{\csisotope}{$^{133}$Cs }
\preprint{APS/123-QED}

\title{Formation of ultracold \potasBnospace \csisotope Feshbach molecules}

 \author{Charly Beulenkamp}%
 \thanks{These authors contributed equally to this work.}
 \author{Krzysztof P. Zamarski}
 \thanks{These authors contributed equally to this work.}
 \affiliation{Universit{\"a}t Innsbruck, Institut f{\"u}r Experimentalphysik und Zentrum f{\"u}r Quantenphysik, Technikerstraße 25, 6020 Innsbruck, Austria
}

 \author{Robert C. Bird}%
 \author{C.~Ruth~{Le~Sueur}}%
 \author{Jeremy M. Hutson}%

\affiliation{Joint Quantum Centre (JQC) Durham-Newcastle, Department of Chemistry, Durham University, South Road, Durham, DH1 3LE, United Kingdom.
}

 \author{Manuele Landini}%
 \author{Hanns-Christoph N\"agerl}%
  \email{christoph.naegerl@uibk.ac.at}
\affiliation{Universit{\"a}t Innsbruck, Institut f{\"u}r Experimentalphysik und Zentrum f{\"u}r Quantenphysik, Technikerstraße 25, 6020 Innsbruck, Austria
}

\date{\today}

\begin{abstract}
We report the creation of an ultracold gas of bosonic \potasBnospace \csisotope molecules. We first demonstrate a cooling strategy relying on sympathetic cooling of \csisotope to produce an ultracold mixture. From this mixture, weakly bound molecules are formed using a Feshbach resonance at 361.7 G. The molecular gas contains $7.6(10)\times 10^3$ molecules with a lifetime of about 130 ms, limited by two-body decay. We perform Feshbach spectroscopy to observe several new interspecies resonances and characterize the bound state used for magnetoassociation. Finally, we fit the combined results to obtain improved K-Cs interaction potentials. This provides a good starting point for the creation of ultracold samples of ground-state \potasBnospace \csisotope molecules.

\end{abstract}
\maketitle

\section{Introduction}
Magnetic-field-tunable Feshbach resonances \cite{FeshbachMoleculeCreationReview} are a powerful tool in experiments on ultracold gases. In addition to offering a simple and reliable way to tune atomic collisions, they serve as a bridge between the high level of control in ultracold atomic gases and the rich physics of molecules. Formation of weakly bound molecules via Feshbach resonances \cite{Cs2Feshbachmolecules,Regal_2003} has led to the observation of molecular BEC \cite{Li2BEC,K2BEC,Li2BEC2} and the study of the BEC-BCS crossover \cite{40KBCS-BEC,6LiBCS-BEC,6LiBCS-BEC2}. Further exploration into ultracold molecule physics was made possible by transferring to deeply bound molecular states by means of stimulated Raman adiabatic passage (STIRAP) \cite{Danzl_2008,KRbgroundstate}. Of particular interest in the last decade have been heteronuclear molecules. These have an electric dipole moment in their ground state, giving rise to long-range dipole-dipole interactions with applications in quantum computing \cite{Theory:quantumcomputationpolarmolecules,Picard_2024}, study of new quantum phases \cite{PhysRevLett.98.060404} and quantum simulation of spin models \cite{Micheli_2006,KRbXYZmodel}, including t-J models \cite{doi:10.1126/science.adq0911} predicted to help explain high-temperature superconductivity \cite{cornish2024quantumcomputationquantumsimulation}. Gases of dipolar molecules have been brought into the degenerate regime, with degenerate Fermi gases created directly from degenerate atomic gases \cite{KRbDegenerateGas, NaKDegenerateMolecules,PhysRevA.107.013307}, and a BEC of bosonic NaCs molecules \cite{NaCsBEC} being realized through evaporative cooling of molecules using microwave shielding of collisions \cite{CaFMicrowaveShielding}.

Ultracold gases of heteronuclear Feshbach molecules have in most cases been created using two different alkali atoms, with over half of the possible combinations realized \cite{KRbFeshbachMol,LiKFeshbachMol,RbCsFeshbachMolecules,NaLiFeshbachMol,Na40KFeshbachMol,Na39KFeshbachMol,NaRbFeshbachMol,NaCsFeshbachMol}. Notably missing from the list of bialkali molecules is KCs, which offers an electric dipole moment of 1.9 D \cite{PhysRevA.109.052814} in its rovibrational ground state and both fermionic and bosonic isotopologues. 
Previous work has predicted Feshbach resonances for different isotopic combinations of K and Cs \cite{Patel:2014} and observed a few of the corresponding atom-loss peaks \cite{GrobnerKCsFeshbach}, but not achieved molecule formation. Here we report the creation of a trapped ultracold sample of \potasBnospace\cs Feshbach molecules by magnetoassociation, providing the necessary starting conditions for the creation of ground-state molecules by STIRAP \cite{KCsSTIRAPpaper}. In Sec.~\ref{sec:cooling} we describe our experimental procedure for preparing an ultracold mixture. In Sec.~\ref{sec:moleculeformation} we discuss the creation and purification of Feshbach molecules. We perform Feshbach spectroscopy (Sec.~\ref{sec:resonancecharacterization}) to observe several new resonances at four different thresholds and characterize the bound state used for magnetoassociation by two-photon bound-state spectroscopy and magnetic-moment measurements (Sec.~\ref{sec:boundstatecharacterization}). Finally, in Sec.~\ref{sec:potentialfitting} we fit the combined results to obtain improved K-Cs interaction potentials that provide a detailed understanding of the collision physics.

\section{Cooling sequence}  \label{sec:cooling} 

Association of atom pairs to form heteronuclear Feshbach molecules requires the preparation of an ultracold mixture of the two species with at least one species close to degeneracy \cite{RbHeteronuclearMolecules}. The collisional properties of \cs and \potasB make them more difficult to cool than other alkalis, and they were among the last alkali isotopes to be brought to degeneracy \cite{39KfirstBEC,CsfirstBEC}. In previous work, we used established techniques to cool the two species sequentially before mixing  \cite{GrobnerKCsFeshbach}. This scheme had the advantage of allowing experiments on either single-species BEC or mixtures. It could in principle be used to produce a mixture at high enough phase-space density for magnetoassociation, but required long cycle times and significant experimental complexity. To simplify our experimental sequence we explored a cooling strategy similar to those used for RbCs \cite{RbCsoverlappingBECs} and NaCs \cite{NaCsBEC}, where the two species are mixed throughout the cooling process and Cs is cooled sympathetically. These experiments rely on radiofrequency (RF) evaporation of the lighter atom in a magnetic trap to load the mixture into an optical dipole trap. This strategy is not viable for the case of \potasB, as its background scattering length of $-35\ a_0$ leads to a Ramsauer-Townsend minimum in the collisional cross-section at collision energies around 400 \unit{\micro\kelvin} \cite{39Kcollisions}. Instead, we use an established technique for preparing \potasB BEC \cite{39KaloneBEC}, with modifications to allow the \potasB gas to cool Cs sympathetically. In the following we describe our cooling scheme, and detail how to manage the sympathetic cooling of Cs.

\begin{figure*}[]
  \centering
  \includegraphics[width=0.9\textwidth]{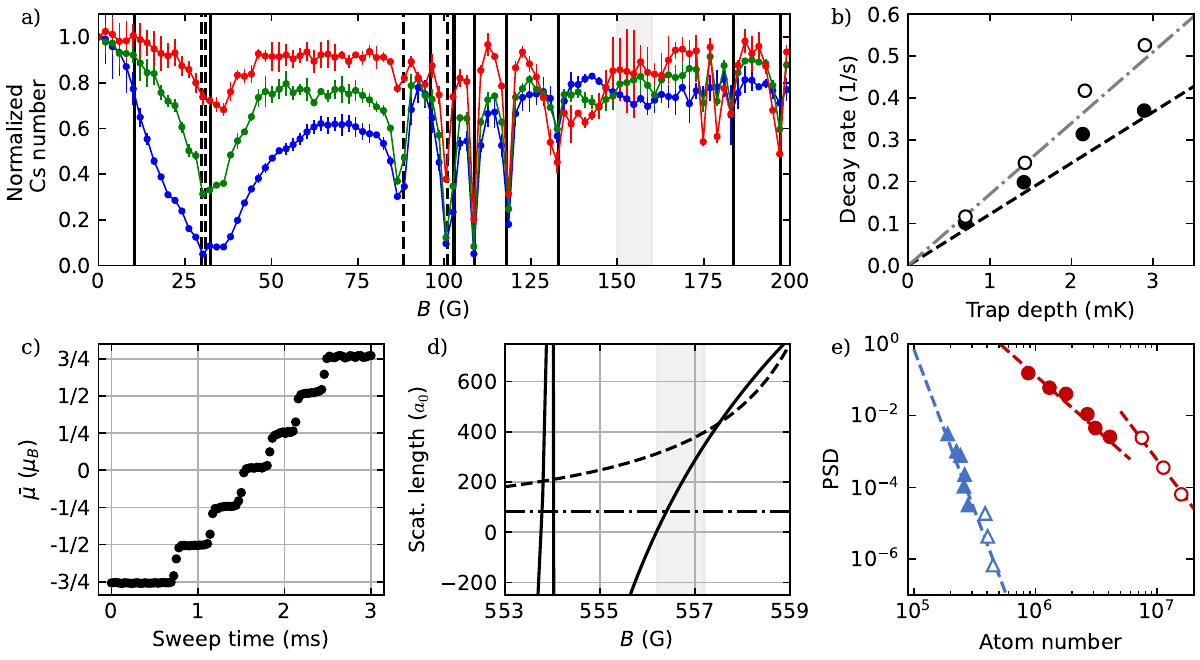}
  \caption{Cooling of the mixture. 
  (a) Loss spectroscopy of Cs in the state $(3,-3)$ . We plot the fraction of remaining Cs atoms after a 1 second hold time at initial peak density $1.5(5)\times 10^{12}$\unit{\cm^{-3}} and temperatures 84 \unit{\micro\kelvin}(red), 43 \unit{\micro\kelvin} (green) and 21 \unit{\micro\kelvin} (blue) as a function of the magnetic field $B$. Error bars are the standard deviations from 3 experimental runs. Solid (dashed) vertical lines mark peaks (zero crossings) in the imaginary (real) part of the scattering length \cite{CsCollisionsTheory}. The shaded area marks a magnetic field range over which the K scattering length can be tuned from 50 $a_0$ to 300 $a_0$ \cite{39KpinpointFeshbach}. (b) Cs decay rate as a function of ODT depth for polarizations perpendicular (filled dots) and parallel (hollow dots) to the magnetic bias field. Error bars are smaller than the plot markers. The dashed (dash-dotted) line is the predicted Raman scattering rate for perpendicular (parallel) polarization. (c) Stern-Gerlach measurements of the average magnetic moment of the Cs cloud during the trap-mediated Raman transfer. Error bars correspond to the uncertainty in the cloud position as determined by fitting a Gaussian distribution to the density distribution, and are smaller than the plot markers. (d) Scattering lengths of  $(1,-1)_{\text{K}}$ (dashed) \cite{39KpinpointFeshbach},  $(3,3)_{\text{Cs}}$ (solid) \cite{CsCollisionsTheory} and $(1,-1)_{\text{K}}$+$(3,3)_{\text{Cs}}$ (dash-dotted) \cite{GrobnerKCsFeshbach} near 557 G, as a function of the magnetic field $B$. The shaded area marks the region where condensates of K and Cs are both miscible and stable. (e) Phase-space density (PSD) as a function of atom number for K (dots) and Cs (triangles) during evaporative cooling in the ODT at 150 G (hollow) and 557 G (filled). The dashed lines indicate evaporation efficiencies of 3 and 5 for K and 9 for Cs.}
  \label{fig:dimpleEvap}
\end{figure*}

Our setup has been previously described in detail in Ref.\ \cite{GrobnerApparatus}. It consists of a steel chamber fed with precooled atoms from separate 2D$^+$ MOTs \cite{PhysRevA.58.3891} of \potasB and \cs  (referred to as K and Cs in the remainder of Sections~\ref{sec:cooling} and \ref{sec:moleculeformation}). We start our experimental sequence by loading a dual-species MOT with typical atom numbers of $3\times 10^8$ for K and $2\times 10^7$ for Cs. After loading, the MOT is compressed by increasing the magnetic field gradient from 9.5 G/cm to 21 G/cm while reducing the repumping intensities. We then shut off the gradient and apply D1-line grey molasses \cite{K39Greymolasses2013} on K and D2-line molasses on Cs to cool to temperatures of 13(1) \unit{\micro\kelvin} and 14.2(2) \unit{\micro\kelvin}, respectively, followed by optical pumping of K to $(f,m_f)=(1,-1)$ \footnote{To describe pairs of atoms, it is necessary to distinguish between quantum numbers for the individual atoms and those for the pair. We adopt the widely used convention of using lower-case letters for the individual atoms and upper-case letters for the pair.}. 
The mixture is then trapped in a quadrupole magnetic trap with an initial magnetic field gradient of 35 G/cm, trapping only atoms in the states $(1,-1)_{\text{K}}$ and $(3,-3)_{\text{Cs}}$. The trap is held at this gradient for 100 ms to ensure spin purity of the trapped gas, followed by a ramp up to 65 G/\unit{\cm}. We then shine a deep optical dipole trap (ODT) (23W, 30 \unit{\micro\meter} waist, 1064 nm) near the center of the magnetic trap. The trap depth for K of $U_{\text{K}} =2.1$ mK ensures that typical collision energies overcome the Ramsauer-Townsend minimum \cite{39KaloneBEC}. After 2.5 seconds of loading, we switch off the magnetic trap. After equilibration the ODT typically contains $1.6(1) \times 10^7$ K atoms and $4.5(1)\times 10^5$ Cs atoms at temperatures of 388(4) \unit{\micro\kelvin} and 394(4) \unit{\micro\kelvin}. 

The mixture can now be further cooled by evaporation near an intraspecies Feshbach resonance for K. The interspecies scattering length takes values between 60 $a_0$ and 80 $a_0$ away from interspecies Feshbach resonances, ensuring sufficiently high elastic collision rates for the two species to thermalize. The ratio of trap depths, $U_{\text{Cs}}/U_{\text{K}} = 1.94$ at 1064 nm, suppresses direct evaporation of Cs. Cs is therefore sympathetically cooled by K. Since we load Cs in the state $(3,-3)_{\text{Cs}}$ into a tight trap, we have to consider collisional loss processes, which can inhibit efficient sympathetic cooling. Dipolar relaxation couples the spins of a colliding atom pair to their relative angular momentum, resulting in heating and depolarization. Additionally, the typically large scattering length of Cs results in high three-body recombination rates. Both two-body and three-body loss rates have an upper bound in the unitary limit $k a \gg 1$, where $k$ is the wavenumber of a colliding pair and $a$ is the scattering length. 
The limiting of the loss rate in the unitary regime can be seen in loss spectroscopy of the Cs cloud at different temperatures (Fig.\ \ref{fig:dimpleEvap}(a)). Loss rates are high near zero-crossings of the scattering length, and become more pronounced at lower temperatures \cite{CsFeshbachTweezers}. At magnetic fields between 150 G and 160 G and temperatures above 10 \unit{\micro\kelvin}, the Cs loss rates remain low enough for efficient sympathetic cooling. The scattering length for K can be tuned using a broad Feshbach resonance at 162 G, making it suitable for the first step of evaporative cooling.

In addition to loss due to Cs-Cs collisions, we observe a one-body decay of the Cs atom number with a rate proportional to the ODT depth. To separate one-body loss from two-body and three-body loss in the Cs cloud, we reduce the density to below $2 \times 10^{11}\ \unit{\cm^{-3}}$ and set the magnetic field to 150~G. Fig.\ \ref{fig:dimpleEvap}(b) shows the decay rate as a function of the trap light intensity averaged over the Cs cloud, expressed in terms of the Cs trap depth. We interpret the trap-depth-dependent decay as depolarization of Cs atoms due to scattering of trap light, followed by exothermic spin-changing collisions. Scattering of trap photons can leave the atom in the same internal state (Rayleigh scattering), or a different one (Raman scattering) \cite{Cline:94}. Laser light with an intensity $I_L$, frequency $\omega_L$ and polarization unit vector $\boldsymbol{u}_L$ induces transitions from an initial state $g$ to a final state $g'$ at a rate that can be expressed in terms of the atomic polarizability as (see Supplemental Material \cite{SuppMat}):
\begin{equation}
\Gamma_{g\rightarrow g'} =  I_L \frac{\omega_L^3}{3 \pi \hbar c^4 \epsilon_0^2 }  \bigg| \langle g'|    \alpha^\textrm{s}_{nJ} \boldsymbol{u}_L - i \alpha^\textrm{v}_{nJ}  \frac{ \boldsymbol{u}_L \times   \hat{\boldsymbol{J}}}{2J}   |g \rangle \bigg|^2 ,
\label{eq:PhotonScatteringRate}
\end{equation}
where $\hat{\boldsymbol{J}}$ is the total electronic angular momentum operator of the atom, while $\alpha^\textrm{s}_{nJ}$ and $\alpha^\textrm{v}_{nJ}$ are the scalar and vector polarizabilities, respectively.  Due to its large fine-structure splitting relative to the trap-light detuning from the D1 and D2 lines, Cs has a significant vector polarizability at 1064 nm of $\alpha^\textrm{v} = -200$ $a_0^3$ \cite{Le_Kien_2013}, which results in a higher Raman scattering rate at this wavelength than for other alkalis. Most Cs atoms that change state end up in levels with $f=4$, after which they can undergo exothermic two-body collisions with K, leading to atom loss. We compute a corresponding loss rate by assuming that every Raman scattering event results in the loss of the atom and find good agreement between calculated and measured loss rates. The depolarizing effect of the trap light will affect not just evaporative cooling, but also the ODT loading process. The ODT depth must be sufficiently high to load K atoms, but a further increase of trap depth will result in fast decay of loaded Cs atoms, limiting the amount of Cs that can be loaded. 
% e^2 a^2_0/E_h

Having characterized the loss processes for Cs, we now proceed with evaporative cooling in the ODT.  We start evaporation at 150 G by reducing the ODT power from 23 W to 1 W in 250 ms. We then hold the cloud for 250 ms at 114 G to remove spin impurities. At this field, impurities in state $(1,0)_{\text{K}}$ (probably originating from spin-relaxation collisions between K atoms during ODT loading) are quickly depleted by three-body recombination due to a Feshbach resonance for the collisions with atoms in $(1,-1)_{\text{K}}$. Cs spin impurities undergo exothermic spin exchange with $(1,-1)_{\text{K}}$ atoms, which ejects them from the trap. 

Beyond this point, evaporative cooling using the $(3,-3)_{\text{Cs}}$ state becomes inefficient due to high two-body and three-body Cs-Cs loss rates, so we proceed in the channel  $(1,-1)_{\text{K}}+(3,3)_{\text{Cs}}$ at 557~G. As there is no way to couple the states $(3,-3)_{\text{Cs}}$ and $(3,3)_{\text{Cs}}$ directly, spin transfer requires briefly populating intermediate states. These intermediate states suffer from much higher collisional loss rates, so time spent in them should be minimized. Achieving high Rabi frequencies using RF coupling is complicated by constraints to antenna geometry and RF pickup in surrounding electronics. Instead, we exploit the large vector polarizability of Cs at 1064 nm and drive Raman transitions using our trapping light. We switch on a second beam, labeled RODT, copropagating with the ODT, with orthogonal polarization and a detuning of 41 MHz. By sweeping the magnetic field up by 10 G in 3 ms, we drive a series of adiabatic passages. The spin transfer is shown in Fig.\ \ref{fig:dimpleEvap}(c). We switch off RODT at various times during the magnetic field sweep and perform Stern-Gerlach separation of the Cs cloud. We fit a Gaussian function to the number density of the cloud to determine its center-of-mass position. From the position of the cloud we extract an average magnetic moment $\bar{\mu}$, which can be seen to increase in steps of one quarter of the Bohr magneton $\mu_\textrm{B}$. 

Following the spin transfer of Cs, we immediately increase the magnetic field to 557~G to avoid interspecies spin-exchange collisions that can occur below 355~G. We proceed with evaporative cooling at this magnetic field, reducing the ODT power from 1 W to 50 mW in 1000 ms. The interspecies and intraspecies scattering lengths in this magnetic-field region are depicted in Fig.\ \ref{fig:dimpleEvap}(d). A Cs Feshbach resonance at 548 G \cite{BerningerCsHighField,CsCollisionsTheory} and K resonance at 561 G \cite{39KpinpointFeshbach} result in a window 1.5~G wide where both species can be efficiently cooled and condensates are miscible. The interspecies scattering length in this window is approximately $83\ a_0$, while the intraspecies scattering length for K is above $350\ a_0$ at the Cs zero crossing. As the evaporation progresses and the mixture gets colder, interspecies thermalization slows down while the three-body loss of K speeds up. Evaporation ramps are optimized empirically by maximizing the number of molecules created by magnetoassociation as discussed below. 

To judge the quality of our evaporative cooling scheme, we measure the phase-space densities (PSD) as a function of atom numbers, as depicted in Fig.\ \ref{fig:dimpleEvap}(e). The efficiency of evaporative cooling can be quantified by the parameter $\gamma = -\frac{d \log \rho}{d \log N}$, where $\rho$ is the phase-space density and $N$ is the atom number. For K we find $\gamma \approx 5$ initially, dropping to $\gamma \approx 3$ at 557 G. For Cs we find $\gamma \approx 9$, which indicates efficient sympathetic cooling.  At the end of evaporation our mixture consists of $ 8.7(3)\times 10^5$ K atoms at a PSD of $0.15(2)$, and $1.9(1) \times 10^5$ Cs atoms at a PSD of $0.0031(4)$.  The clouds remain well overlapped, as the estimated differential gravitational sag of $1.4$ \unit{\micro\meter} is small compared to the computed 1/$e$ cloud radii of $9.1$ \unit{\micro\meter} and $7.6$ \unit{\micro\meter} for K and Cs, respectively. 

In an unbalanced mixture, the molecule creation efficiency $N_\text{mol}/N_\text{atom}$ with respect to the minority component depends primarily on the phase-space density of the majority component \cite{RbHeteronuclearMolecules}. Cooling K closer to degeneracy would therefore increase the molecule creation efficiency with respect to Cs. However, evaporative cooling beyond the point described in the previous paragraph becomes very inefficient for Cs, because a tilt of the trap potential due to gravity lowers the trap depth for Cs below that of K. The loss of Cs atoms results in a lower molecule number, negating the gains in molecule creation efficiency from the increased phase-space density of K.

\begin{figure}
  \centering
  \includegraphics[width=0.49\textwidth]{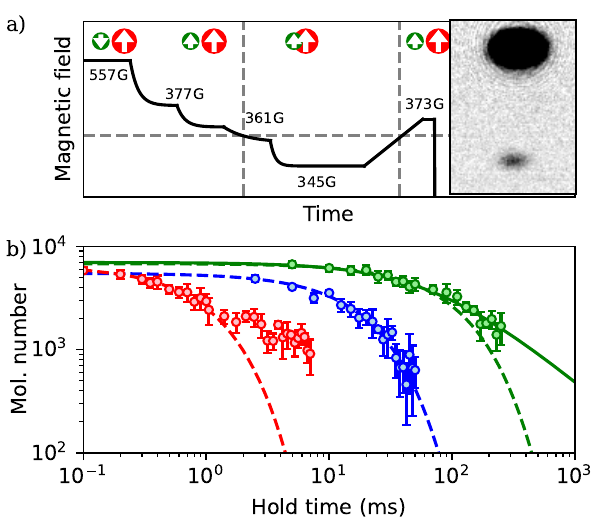}
   \caption{Molecule creation and lifetimes. (a) Schematic of the magnetic-field changes during molecule creation. The horizontal dashed line marks the pole of the Feshbach resonance. The inset on the right shows a typical Stern-Gerlach absorption image of the molecular cloud (below) and remaining K atoms (above). (b) Molecular lifetimes in the presence of both K and Cs (red), only Cs (blue), and for a purified molecular cloud (green). Dashed lines are exponential fits to short hold times. The solid green line is a fit to a two-body decay curve.}
  \label{fig:molecules}
\end{figure}

\section{Molecule formation}\label{sec:moleculeformation}

We now proceed to the creation of weakly bound molecules. Previous experiments have observed six Feshbach resonances between K atoms in states with $f=1$ and Cs atoms in the state $(3,3)_{\text{Cs}}$ \cite{GrobnerKCsFeshbach}. Of these, the most promising for molecule creation is the resonance at 361.7~G in the channel $(1,1)_{\text{K}}+(3,3)_{\text{Cs}}$. This resonance has the largest predicted width, and as both species are in their hyperfine ground states, molecular states below this threshold do not have internal decay channels. To avoid heating and atom loss due to the large background scattering length of Cs, we have to minimize the time between the end of evaporative cooling and molecule creation. The magnetic-field changes during molecule formation are depicted schematically in Fig.\ \ref{fig:molecules}(a). First we jump our magnetic field to 377 G in order to avoid a broad Feshbach resonance at 395 G for atoms in state $(1,1)_{\text{K}}$. The K atoms are then transferred from $(1,-1)_{\text{K}}$ to $(1,1)_{\text{K}}$ in 2 ms by two Raman adiabatic passages. We then jump our magnetic field down close to the interspecies Feshbach resonance, before crossing it at a reduced speed of 3~G/s to ensure adiabatic molecule formation, followed by a jump to 345~G.  After crossing the resonance we apply a magnetic-field gradient of 45~G/cm, which levitates the Feshbach molecules while overlevitating K and Cs atoms. K atoms are ejected from the trap, while Cs atoms remain trapped. To remove the Cs atoms, we ramp down the ODT power from 50 mW to 8 mW in 3 ms. After 10 ms of magnetic separation, the ODT power is ramped back up to 30 mW in 10 ms. This leaves a purified cloud of $7.6(10) \times 10^3$ Feshbach molecules at a temperature of 0.69(5)~ \unit{\micro K} and a PSD of $2.8(8)\times 10^{-4}$. To image the molecules, we dissociate them by ramping the bias field up across the resonance in 2 ms, followed by absorption imaging at zero field of either the K or Cs atoms. 

The molecules can be lost by relaxation to deeply bound states in collisions with either atoms or molecules. After molecule formation, and before the unbound atoms are ejected from the trap, molecules can collide with either K or Cs atoms. We add a delay between molecule creation and magnetic-gradient switch-on and measure the decay of the molecular sample (Fig.\ \ref{fig:molecules}(b)). During the first millisecond, we measure a fast decay with a $1/e$ time of $0.96(4)$ \unit{ms}. This fast decay leads to the loss of a significant fraction of the molecules on the timescale in which K can be ejected from the trap. To confirm this, we switch off the trap directly after magnetoassociation, which reduces atom density. We then observe typical molecule numbers of $11 \times 10^3$,  corresponding to molecule creation efficiencies of 1.4\%(6\%) on K(Cs). 
With K atoms removed but Cs still kept in the trap, the molecule lifetime is increased to $18(2)$ \unit{ms}. This is long enough to allow our magnetic field to stabilize for STIRAP transfer to the ground state, and thus allows the creation of a mixture of Cs atoms and KCs ground-state molecules. Such a mixture is chemically stable under collisions \cite{AlkaliDimerReactions}, and may possess Feshbach resonances \cite{NaKmoleculeatomresonances}. When both K and Cs are removed, the molecular decay is well described by a two-body decay curve $N(t) = N_0/(1+ \Gamma_2 \bar{n_0} t)$ with a loss coefficient of $\Gamma_2 = 5.6(12)\times 10^{-10}$ \unit{\cm^3 s^{-1}}, where $\bar{n_0}$ is the mean initial density. The corresponding $1/e$ lifetime is about 130 ms.

\section{Resonance characterization} \label{sec:resonancecharacterization}

The creation of our ultracold mixture and molecules makes it possible to collect new data on the scattering and bound-state properties of \potasBnospace\csisotope and refit the interaction potentials. Previous Feshbach spectroscopy on a \potasBnospace+\csisotope mixture was carried out by Gr{\"o}bner \textit{et al.} \cite{GrobnerKCsFeshbach}. They observed the positions of six resonances in three different collision channels. They used these to fit interaction potentials based on the singlet and triplet potential curves of Ferber \emph{et al.} \cite{Ferber:2013}. Gr{\"o}bner \textit{et al.} found that the resonances they observed were insufficient to determine both singlet and triplet potentials and could be fitted satisfactorily by modifying only the triplet potential.

\begin{figure*}[t]
\centering
  \includegraphics[width=0.9\textwidth]{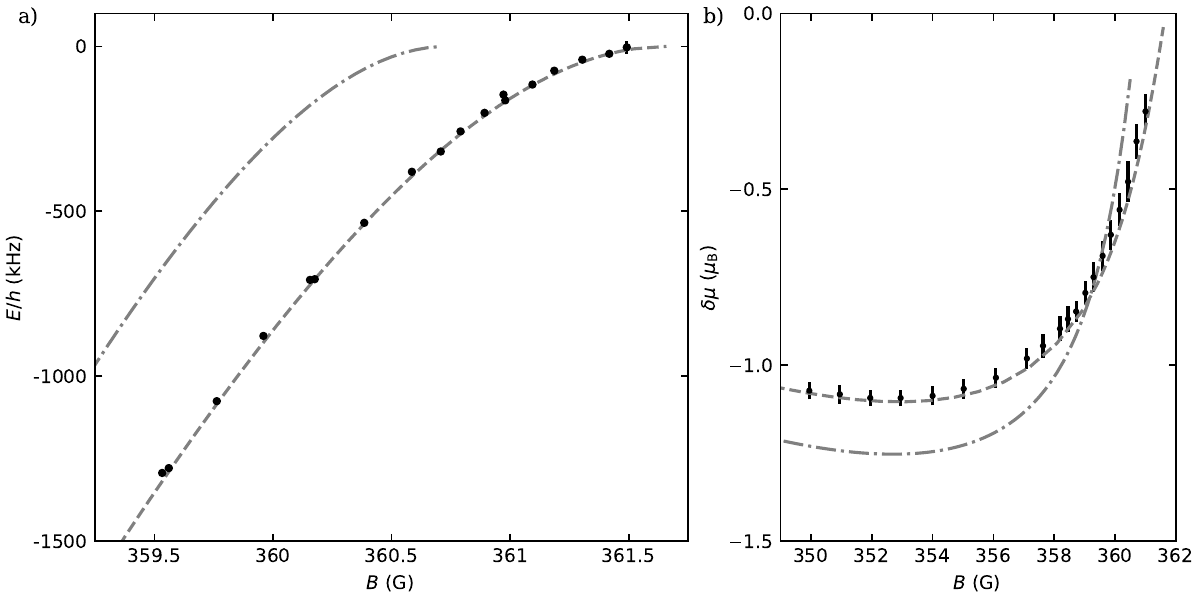}
  \caption{Measurements to characterize the bound state that causes the Feshbach resonance at 361.7 G in the lowest channel. Coupled-channel calculations are plotted for the potentials from Gr{\"o}bner \textit{et al.}\ \cite{GrobnerKCsFeshbach} (dash-dotted) and our refitted potentials (dashed). (a) Binding energies obtained through two-photon spectroscopy as a function of magnetic field $B$. Error bars are smaller than plot markers. The two data points used for refitting the potentials are marked with red diamonds. (b) Differential magnetic moment of the molecules with respect to an atom pair at the $(1,1)_{\text{K}}+(3,3)_{\text{Cs}}$ threshold as a function of magnetic field $B$.}
  \label{fig:BindingEnergy}
\end{figure*}

 To understand and overcome this limitation, we carry out coupled-channel calculations on the bound states responsible for potentially observable resonances in 14 different incoming channels with $M_F=m_{f,\textrm{K}}+m_{f,\textrm{Cs}} \ge 0$. These calculations use the BOUND package \cite{bound+field:2019, mbf-github:2023} with the interaction potentials of Gr{\"o}bner \textit{et al.}\ \cite{GrobnerKCsFeshbach}. The results are shown in Supplementary Material \cite{SuppMat}. In addition to the bound-state energies, we calculate the singlet fraction for each bound state at a magnetic field well removed from any avoided crossing that would perturb its character. The calculated singlet fractions range from 0.088 to 0.479, but those for the resonances observed by Gr{\"o}bner \textit{et al.}\ cover the much smaller range from 0.360 to 0.479. This explains why the singlet and triplet potentials could not be modified independently in ref.\ \cite{GrobnerKCsFeshbach}.

In the present work, we use these calculations to identify resonances with a wide range of singlet fractions that have sufficient strength to be observed by atom loss spectroscopy on a non-degenerate mixture. 
After cooling the ultracold mixture, we drive Raman adiabatic passages to prepare K and Cs in the target spin state and hold the mixture for 50 to 100 ms, after which we measure the number of remaining Cs atoms. In the vicinity of a Feshbach resonance, the mixture will undergo enhanced loss due to three-body recombination. For each resonance, we fit a Lorentzian function to obtain the position of maximal loss. 

\begin{table}[t]
\caption{\label{tab:table1} Overview of newly observed loss peaks observed in Feshbach spectroscopy and comparison with Feshbach resonance positions calculated with the interaction potentials of ref.\ \cite{GrobnerKCsFeshbach}.}
\begin{ruledtabular}
\begin{tabular}{cccc}
\textrm{Threshold}& $(f,m_f)_\textrm{K}+(f,m_f)_\textrm{Cs}$&
\textrm{$B^\text{exp}_\text{loss}$ (G)}&
\textrm{$B^\text{calc,2017}_\text{res}$} \\
\noalign{\vskip 0.5mm} 
\colrule
b+b & $(1, 0)_\text{K}+(3,2)_\text{Cs}$ & 315.551(2)  & 319.35 \\
c+a & $(1,-1)_\text{K}+(3,3)_\text{Cs}$  & 463.251(5) &  none  \\
c+a & $(1,-1)_\text{K}+(3,3)_\text{Cs}$  & 466.230(5) & 467.70 \\
b+c & $(1, 0)_\text{K}+(3,1)_\text{Cs}$  &  299.562(4) & 304.62 \\
d+a & $(2,-2)_\text{K}+(3,3)_\text{Cs}$  &  478.028(4) & 478.85 \\
d+b & $(2,-2)_\text{K}+(3,2)_\text{Cs}$  & 464.579(3)  & 466.01
\end{tabular}
\end{ruledtabular}
\label{table:feshbachspectroscopy}
\end{table}

When discussing resonances at multiple thresholds, explicit quantum numbers become cumbersome. In the remaining sections, we therefore specify atomic states with Roman sequence letters in order of increasing energy. A threshold is identified with a pair of letters, first for K and then for Cs.

The newly observed loss features are shown in Table \ref{table:feshbachspectroscopy} and compared with resonance positions calculated using the interaction potentials of Gr\"obner \emph{et al.}\ \cite{GrobnerKCsFeshbach}. There are discrepancies of up to 5~G in some cases, particularly for the resonances at the b+c and b+b thresholds, which are the ones for which the corresponding bound states have the lowest singlet fractions (0.135 and 0.191, respectively). This justifies refitting the model to optimize the singlet and triplet potentials simultaneously.

\section{Bound-state characterization} \label{sec:boundstatecharacterization}

To provide additional input for refitting the model, we characterize the bound state corresponding to the Feshbach resonance at 361.7 G by measuring its binding energy and magnetic moment. Our binding energy measurements start from the ultracold mixture at the end of evaporative cooling. We first transfer Cs to $(3,2)_\text{Cs}$ and then measure the free-bound transition frequency by two-photon spectroscopy. When the two-photon detuning matches the sum of Zeeman splitting, binding energy and relative kinetic energy of a colliding atom pair, the atomic pair is coupled into the weakly bound molecular state. Transfer to the molecular state shows up as atom loss, as molecules are quickly lost through atom-dimer relaxation or laser light absorption. 

To extend the characterization to magnetic fields further below the Feshbach resonance, we measure the magnetic moment of our purified molecular sample $\mu_{\text{mol}}$ as a function of the magnetic field. After ramping the magnetic bias and gradient fields to their target values, the ODT power is reduced to 5 mW. When the magnetic-field gradient matches the levitation gradient $B'_{\text{lev}} = m_{\text{KCs}}  g/\mu_{\text{mol}}$, molecules remain trapped. We plot the results of the binding-energy and magnetic-moment measurements in Fig.\ \ref{fig:BindingEnergy}(a) and Fig.\ \ref{fig:BindingEnergy}(b), respectively. 

\section{Fitting potential curves}\label{sec:potentialfitting}

The interaction potential curves $V_S(R)$ of refs.\ \cite{Ferber:2013} and \cite{GrobnerKCsFeshbach} are made up of three segments. For each electronic state (singlet X$^1\Sigma^+$, $S=0$, or triplet a$^3\Sigma^+$, $S=1$), the central segment $V_S^\textrm{mid}(R)$ (around the potential minimum) is represented as a high-order power series in a transformed internuclear distance $R$. This is supplemented by a long-range extrapolation based on dispersion coefficients and a short-range extrapolation based on a simple inverse-power expansion. The central segment was fitted to extensive electronic spectra from Fourier-transform spectroscopy in ref.\ \cite{Ferber:2013}. The details of the potential functions are given in refs.\ \cite{Ferber:2013} and \cite{GrobnerKCsFeshbach}.

The physical quantities that principally determine the near-threshold bound states and the positions of Feshbach resonances are the singlet and triplet scattering lengths, $a_\textrm{s}$ and $a_\textrm{t}$, and the leading dispersion coefficient $C_6$. The singlet and triplet scattering lengths depend on the entire potentials $V_S(R)$. In the present work, we wish to adjust $a_\textrm{s}$ and $a_\textrm{t}$ to reproduce the observed ultracold properties, while retaining the fit to the Fourier-transform spectra as much as possible. We do this by retaining the power-series expansion coefficients for $V_S^\textrm{mid}(R)$ from ref.\ \cite{Ferber:2013}, while adjusting the short-range extrapolations. The adjustments we make have only small effects on the energies of deeply bound vibrational states.

An important aspect of fitting interaction potentials is to ensure that parts of the potential that are \emph{not} well determined by the experiments are constrained to have physically reasonable behaviour.
Previous work on potential curves for the alkali dimers has used short-range extrapolations of the form
\begin{equation}
V_S^\textrm{SR}(R) = A'_S + B'_S/R^{N_S}  \qquad \hbox{for } R < R_S^\textrm{SR}.
\label{eq:sr-power}
\end{equation}
$A'_S$ and $B'_S$ are usually chosen to match the value and derivative of $V_S^\textrm{mid}(R)$ at $R_S^\textrm{SR}$. However, the values of the arbitrary powers $N_S$ have ranged from 1.8 \cite{Zhu:NaK:2017} to 12 \cite{Tiemann:2020}, producing unphysical variations in the hardness of the repulsive wall between systems. In the present work, we replace Eq.\ \ref{eq:sr-power} with
\begin{equation}
V_S^\textrm{SR}(R) = A_S + B_S\exp(-\alpha_S R)  \qquad \hbox{for } R < R_S^\textrm{SR}.
\label{eq:sr-exp}
\end{equation}
Ihm \emph{et al.}\ \cite{Ihm:1990} have shown that, for a variety of systems, the repulsive part of the interaction between two atoms is well represented by an exponent $\alpha\approx 0.85\beta$, where $\beta$ is the average of $\sqrt{8E_\textrm{I}(m_e/\hbar^2)}$ for the two atoms and $E_\textrm{I}$ is the ionization energy. We have found that this approach gives a good representation of the short-range potentials for pairs of alkali-metal atoms \cite{Crome:2021}. For KCs, $0.85\beta = 1.767$~\AA$^{-1}$.
$A_S$ and $B_S$ are again chosen to match the value and derivative of the mid-range potential at $R_S^\textrm{SR}$. In fitting with this functional form for KCs, we choose values of $R_0^\textrm{SR}=3.5$~\AA\ and $R_1^\textrm{SR}=5.25$~\AA\ that allow sufficient variations of scattering length for small variations of $\alpha_S$ around $0.85\beta$. We then vary $\alpha_S$, choosing $A_S$ and $B_S$ for each $\alpha_S$ to match the value and derivative of $V_S^\textrm{mid}(R)$ at $R_S^\textrm{SR}$.

Five of the six new resonances observed here are due to s-wave bound states crossing threshold. The exception is the feature near 463~G at the c+a threshold, which cannot be explained as an s-wave feature and is considered further below. We carry out least-squares fits of $\alpha_0$ and $\alpha_1$ to these five resonances, together with the six resonances observed by Gr\"obner \emph{et al.} \cite{GrobnerKCsFeshbach}. We also include two data points to encapsulate the measurements of binding energies $E_\textrm{b}$ described above: these are the fields at which $E_\textrm{b}/h$ is 10.7 and 1281.5 kHz. Including both these ensures that both the position of the bound state and its curvature close to threshold are accurately reproduced; the curvature is a direct reflection of the width of the resonance, which is sensitive to $a_\textrm{s}-a_\textrm{t}$.

\begin{table}[tbp]
\caption{The parameters of the fitted short-range potential of Eq.\ \ref{eq:sr-exp}. The quantities in parentheses are standard errors including parameter correlation. All digits quoted are needed to reproduce the observables accurately in calculations. The derived parameters $A_S$ and $B_S$ are included for convenience. The complete potential curves are obtained by combining these with the mid-range and long-range potentials of ref.\ \cite{Ferber:2013}.
\label{tab:pot-params}} \centering
\begin{ruledtabular}
\begin{tabular}{lccl}
Parameter & singlet & triplet & unit \\
\hline
\noalign{\vskip 1mm} % to avoid the superscript touching the hline  
$R_S^\textrm{SR}$ & 3.5                   & 5.25                  & \AA \\
$\alpha_S$        & 1.70342(235)          & 1.726(15)             & \AA$^{-1}$ \\
$A_S/hc$          & $-5402.8994$           & $-388.7871$            & cm$^{-1}$ \\
$B_S/hc$          & $1.224302 \times 10^6$ & $2.266474 \times 10^6$ & cm$^{-1}$ \\
$a_S$             & $-29.2(8)$            & 77.7(2)               & $a_0$
\end{tabular}
\end{ruledtabular}
\end{table}

Our least-squares fits use the interactive fitting package I-NoLLS \cite{I-NoLLS}, which allows us to include multiple assignments of observables and adjust them manually as the fit proceeds. During the fitting process, all the properties are calculated using the FIELD package \cite{bound+field:2019, mbf-github:2023}. The fitted potential parameters are given in Table \ref{tab:pot-params}, including the derived quantities $A_S$ and $B_S$ of Eq.\ \ref{eq:sr-exp}. The quality of fit to the observables is given in Table \ref{tab:fit-quality}. The resonance positions are fitted to within 1~G, except for the one at 599.32 G, whose interpretation is complicated by an avoided crossing very close to threshold. This level of agreement is reasonable in view of the complicated relationship between the positions of resonances and loss peaks \cite{Wang:3-body:2014}.

\begin{table}[tbp]
\caption{Quality of fit to resonance positions and bound-state energies, together with the singlet fractions of the bound states with vibrational quantum number $n=-2$ that mostly determine the resonance positions. The quantities listed as ``Unc." determine the relative weights used in the fits; they are composite quantities based on physical arguments including model dependence, not experimental uncertainties.}
\label{tab:fit-quality} \centering
\begin{ruledtabular}
\begin{tabular}{clclc}
Threshold & $B_\textrm{loss}^\textrm{exp}$ (G) & Unc.\ (G) & $B_\textrm{res}^\textrm{calc}$ (G) & Singlet fraction\\
\hline
a+a & 442.59 & 0.4 & 443.45 & 0.482  \\
b+a & 513.12 & 0.3 & 513.77 & 0.478  \\
b+a & 419.30 & 0.3 & 420.24 & 0.360  \\
b+b & 315.57 & 0.2 & 315.77 & 0.190  \\
c+a & 466.24 & 0.2 & 465.61 & $n=-1$ \\
c+a & 491.50 & 0.4 & 492.24 & 0.394  \\
c+a & 599.32 & 0.3 & 597.78 & 0.479  \\
b+c & 299.57 & 0.2 & 299.87 & 0.135  \\
d+a & 478.05 & 0.2 & 478.07 & 0.306  \\
d+b & 464.60 & 0.2 & 464.62 & 0.239  \\

\hline
$E_\textrm{b}/h$ (kHz) & $B^\textrm{exp}$ (G) & Unc.\ (G) & $B^\textrm{calc}$ (G) & \\
\hline
  10.7 & 361.4995 & 0.03 & 361.4901 & \\
1281.5 & 359.5660 & 0.02 & 359.5668 & \\
\end{tabular}
\end{ruledtabular}
\end{table}

It is conceptually useful to interpret the fitted potentials in terms of the resulting singlet and triplet scattering lengths, which are $a_\textrm{s}=-29.2(8)\ a_0$ and $a_\textrm{t}=77.7(2)\ a_0$, respectively. The value of $a_\textrm{t}$ for the current potential is outside the error bounds of the value $a_\textrm{t}=74.88(9)\ a_0$ obtained in ref.\ \cite{GrobnerKCsFeshbach}. This apparent discrepancy arises because $a_\textrm{s}$ and $a_\textrm{t}$ are correlated. The effect of this correlation on uncertainties was implicitly neglected in ref.\ \cite{GrobnerKCsFeshbach}, where $a_\textrm{s}$ was fixed at the value from ref.\ \cite{Ferber:2013}. Our ability to determine $a_\textrm{s}$ and $a_\textrm{t}$ independently here comes both from the inclusion of resonances with a wider range of singlet fractions than in ref.\ \cite{GrobnerKCsFeshbach} and from including precise measurements of binding energies. As mentioned above, the curvature of the binding energy near threshold is sensitive to $a_\textrm{s}-a_\textrm{t}$; this is significantly larger for the current potential than for the potential of ref.\ \cite{GrobnerKCsFeshbach}, and it may be seen in Fig.\ \ref{fig:BindingEnergy} that both the curvature near threshold and the magnetic moment well below threshold are much better represented by the current potential.

\begin{figure}[t]
\begin{center}
\includegraphics[width=0.46\textwidth]{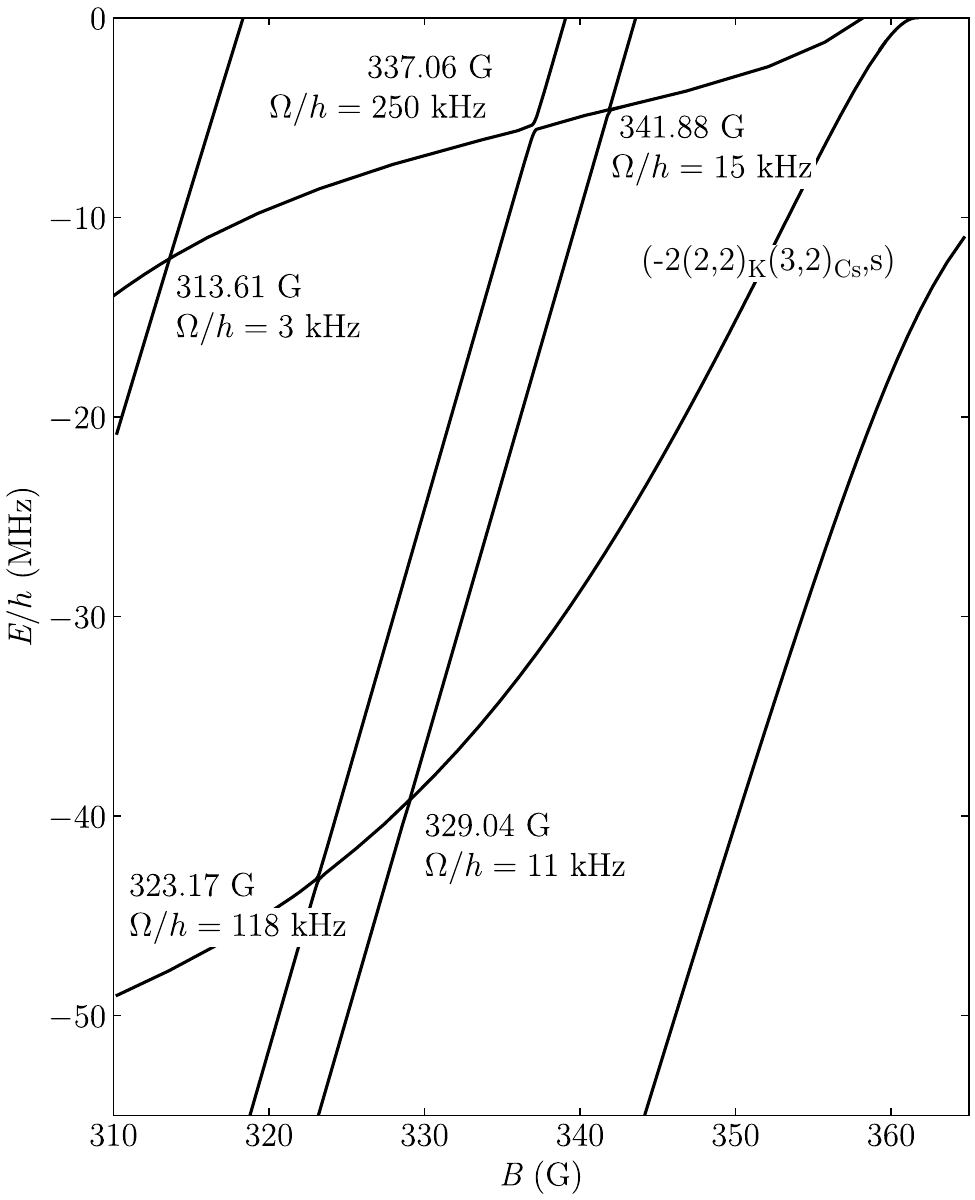}
\caption{Bound states of $^{39}$KCs with $M_F=4$ as a function of magnetic field $B$, shown relative to the lowest threshold. Avoided crossings are labeled with the field and the effective coupling matrix element $\Omega$ between the states. The state that is observed experimentally is labeled with its quantum numbers $(n(f,m_f)_\textrm{K}\,(f,m_f)_\textrm{Cs},L)$.} 
\label{fig:L2-states}
\end{center}
\end{figure}

We have used the new interaction potentials to calculate bound-state and scattering properties of $^{39}$K+Cs. The scattering calculations used the MOLSCAT package \cite{molscat:2019, mbf-github:2023}. We obtained resonance positions, widths and background scattering lengths \cite{Frye:resonance:2017} for all the broad s-wave resonances that occur at the lowest threshold for each $M_F\ge 0$ at fields below 600~G. They are listed in the Supplementary Material \cite{SuppMat}. In addition, Fig.\ \ref{fig:L2-states} shows the near-threshold bound states below the lowest threshold between 310 and 370 G, including d-wave states (with $L=2$). The s-wave state that produces the resonance near 361.7~G has character $(n(f,m_f)_\textrm{K})(f,m_f)_\textrm{Cs},L) = (-2(2,2)_\textrm{K}(3,2)_\textrm{Cs},\textrm{s})$ at depths less than 30~MHz below threshold, but has a broad avoided crossing with $(-1(1,1)_\textrm{K}(3,3)_\textrm{Cs},\textrm{s})$ below that.  It undergoes narrow avoided crossings at 323.17 and 329.04 G with d-wave states. As described in Supplementary Material \cite{SuppMat}, we observe losses of molecules in magnetic field ramps across 329.21(2) G, which we interpret as nonadiabatic processes due to the d-wave state. This confirms that the present potential works well for d-wave as well as s-wave states. There is an additional, shallower, d-wave state that crosses threshold near 358 G and has a strong avoided crossing with another d-wave state that appears at the lower right of the figure. The improved potentials can be used to compute accurate bound-state wavefunctions. With the addition of excited-state wavefunctions, we will be able to construct a model for STIRAP intensities \cite{10.21468/SciPostPhys.15.6.220} and identify optimal initial states for transfer to the hyperfine ground state. 

We made an attempt to determine the dispersion coefficient $C_6$ from the present experimental data. This is potentially possible because most of the resonances observed here are due to bound states with vibrational quantum number $n=-2$ with respect to threshold, while the resonance near 466~G at the c+a threshold is due to a state with $n=-1$. The spacing between states with different values of $n$ depends sensitively on $C_6$ \cite{Gao:2000}. However, by contrast with RbCs \cite{Takekoshi:RbCs:2012} and NaCs \cite{Brookes:2022}, we were unable to determine a value of $C_6$ for KCs that differs significantly from the theoretical value \cite{Derevianko:2001}. Reducing the uncertainty in $C_6$  would require more extensive spectroscopic data or measurements of resonances involving states with a wider range of vibrational quantum numbers.

The interaction potentials do not provide a good explanation of the experimental loss feature observed at 463.27 G at the c+a threshold. It cannot be due to an s-wave resonance, and even the nearest d-wave resonances lies at 459.1(3) G on the present fitted potential. The loss peak might arise from a mechanism other than a Feshbach resonance, such as a laser-driven process \cite{Cho:RbCs:2013}.

\section{Conclusions}
We have demonstrated a method for producing ultracold mixtures of \potasB and \cs. Th method relies on sympathetic cooling of Cs by $^{39}\text{K}$ to reach sufficiently high phase-space densities for molecule creation. With minor modifications, our technique should be able to produce similar mixtures of \potasBB and $^{133}\text{Cs}$. Further improvements to the cooling scheme may allow the creation of overlapping BECs. The high phase-space density  of BECs can be exploited to form molecules at high filling fractions from dual Mott insulators in an optical lattice \cite{RbCsLattice,KRblattice}. 

From our ultracold mixture, we can produce $7.6\times 10^3$ \potasBnospace \csisotope  Feshbach molecules by magnetoassociation using a Feshbach resonance at 361.7~G. Our purified molecular samples have lifetimes of about 130 ms, limited by inelastic two-body collisions between molecules. They provide an excellent starting point for the creation of ultracold ground-state \potasBnospace \csisotope  molecules using STIRAP transfer \cite{KCsSTIRAPpaper}

We have used coupled-channel calculations to analyse the Feshbach resonances used to fit interaction potentials in previous work. We found that the bound states responsible for the resonances span a quite limited range of singlet fractions. Guided by coupled-channel calculations, we have measured the positions of resonances with a much wider range of singlet fractions. In addition, we have performed measurements of binding energies for the bound state corresponding to the 361.7~G resonance. We have used the measurements on bound states and resonance properties to determine new interaction potentials for K+Cs. The additional measurements in the present paper made it possible to fit the singlet and triplet interaction potentials independently and to obtain well-determined singlet and triplet scattering lengths. The new potentials give a very satisfactory picture of the ultracold properties of $^{39}$K+Cs. They could in future be used to make predictions for other isotopic combinations, using coupled-channel calculations with appropriate atomic masses and hyperfine coupling constants.

\begin{acknowledgments}
We thank Govind Unnikrishnan and Dechao Zhang for technical contributions at an early stage of the experiment. 

We acknowledge funding from the European Research Council (ERC) under Project No.\ 789017, the FWF under Project No.\ P29602-N36, a Wittgenstein prize grant under FWF Project No.\ Z336-N36, an FFG infrastructure grant with project number FO999896041, the FWF’s COE 1 and quantA, and the Engineering and Physical Sciences Research Council (EPSRC) under Grant Nos.\ EP/P01058X/1, EP/T518001/1, and EP/W00299X/1. K.P.Z. acknowledges support from the Austrian Science Fund (FWF) within the DK-ALM (Grant No.\ W1259-N27).
This research was funded in whole or in part by the Austrian Science Fund (FWF) Grant DOI 10.55776/W1259. For open-access purposes, the authors have applied a CC BY public copyright license to any author accepted manuscript version arising from this submission. 

CB and KPZ developed the experimental procedures, performed the measurements and carried out the data analysis. ML and HCN supervised and advised on the experimental work. RCB, CRLS and JMH carried out coupled-channel calculations and fitted the interaction potentials.

Data supporting this study is publicly available from Zenodo at \href{https://zenodo.org/records/15478801}{10.5281/zenodo.15478800}.

\end{acknowledgments}

\nocite{PhysRevA.61.012501,Hutson:res:2007,Hutson:CPC:1994,Hutson:expect:88,Manolopoulos:1986,Alexander:1987,Brookes:2022} %References from the supplementary materials

\bibliography{main}% Produces the bibliography via BibTeX.

\hyphenation{Fesh-bach}
\hyphenation{bi-alkali}

\preprint{APS/123-QED}
\onecolumngrid

\clearpage

\supplementarysection

%\begin{bibunit}
\subsection{Spontaneous Raman Scattering}

We provide a derivation of Eq. 1, which is a way of writing the Kramers-Heisenberg formula \cite{Cline:94} at detuning from resonance in terms of the atomic polarizability. We start with the semiclassical Stark operator, which describes the interaction between the atom and the electromagnetic field with complex amplitude $\boldsymbol{\mathcal{E}}$ \cite{Le_Kien_2013}:
\begin{equation}
    \hat{V}^S = - \frac{1}{4} \bigg\{ \alpha^\textrm{s}_{nJ} (\omega)  \boldsymbol{\mathcal{E}}^* \cdot \boldsymbol{\mathcal{E}}- i \alpha^\textrm{v}_{nJ}(\omega)   \frac{ [  \boldsymbol{\mathcal{E}}^* \times \boldsymbol{\mathcal{E}}]   \cdot    \hat{\boldsymbol{J}}}{2J}  \bigg\}, \label{eq:StarkOperator}
\end{equation}
where the tensor polarizability is omitted, as it is equal to zero for an atom with $J = 1/2$. Note that this operator as written here is not well defined when the electric field contains multiple frequency components. However, when applying Fermi's golden rule, energy conservation requires that the frequency of the scattered photon differs from that of the laser photon by no more than the hyperfine and Zeeman splittings, which are negligible compared to the detuning in our case. We can thus proceed using the values of the polarizability at the laser frequency and drop the frequency argument from the notation.

To treat the vacuum modes, we make use of second quantization of the electromagnetic field. We start with the electric field modes in a cubic box of volume $V$, and replace the classical field $\boldsymbol{\mathcal{E}}$ with the operator
\begin{equation}
\hat{\boldsymbol{\mathcal{E}}} = \sum_{\boldsymbol{k},\mu}  E_{V,\boldsymbol{k}} \boldsymbol{u}_{\boldsymbol{k},\mu} \hat{a}_{\boldsymbol{k},\mu} e^{i \boldsymbol{k}\cdot \boldsymbol{r} } +    E_\text{L,1} \boldsymbol{u}_L \hat{a}_{L} e^{i \boldsymbol{k}_L\cdot \boldsymbol{r}},
\end{equation}
Where $E_{V,\boldsymbol{k}} = \sqrt{\frac{\hbar c |\boldsymbol{k}|}{2 V \epsilon_0}}$,  $\boldsymbol{u}_{\boldsymbol{k},\pm}$ are the polarization vectors corresponding to the two possible projections of the photon spin along $\boldsymbol{k}$, and $\hat{a}_{\boldsymbol{k},\mu}$  and $\hat{a}^\dagger_{\boldsymbol{k},\mu}$ are annihilation and creation operators of modes with wavenumber $\boldsymbol{k}$ and polarization $\mu$, respectively. The mode of the trapping laser is separated for convenience. 

The laser light is best described by a coherent state. In the limit of high intensities this state is sharply peaked in the Fock basis. We can treat states with $n_L$ and $n_L-1$ photons as equivalent, and identify the expectation value of the electric field with the classical amplitude $E_L$:
\begin{equation}
    \bra{n_L-1} \hat{\boldsymbol{\mathcal{E}}} \ket{n_L} \approx  E_\text{L,1} \boldsymbol{u}_L \sqrt{n_L} e^{i \boldsymbol{k}_L\cdot \boldsymbol{r}} \approx  E_\text{L} \boldsymbol{u}_L  e^{i \boldsymbol{k}_L\cdot \boldsymbol{r}},
\end{equation}
and we substitute $E_\text{L,1} \sqrt{n_L} = E_L$ later on.

We consider states of the form $|g,n_V, n_L \rangle$, where $g$ is the atomic state, $n_V$ is the number of photons in a certain vacuum mode, and $n_L$ is the number of photons in the laser mode. All other vacuum modes are assumed to have zero occupation. We ignore the motion of the atom, and drop the $e^{i \boldsymbol{k}\cdot \boldsymbol{r}}$ terms in the electric field, which couple different momentum states of the atom. We want to know the matrix elements corresponding to transfer of a photon between the laser mode and a vacuum mode

\begin{align}
   V_{g',g} = &\langle g',1,n_L-1 | \hat{V}^S | g',0,n_L \rangle \\
  = &-\frac{1}{4} \langle g', 1, n_L-1 | \bigg[ \alpha^\textrm{s}_{nL} (\boldsymbol{u}_{k,\pm}^* \cdot \boldsymbol{u}_L ) - i \alpha^\textrm{v}_{nL}  \frac{ [ \boldsymbol{u}_{k,\pm}^* \times \boldsymbol{u}_L ]   \cdot \hat{\boldsymbol{J}}}{2J}  \bigg] E_{V,k}^* \hat{a}^\dagger_{k,\mu} \hat{a}_{L} E_\text{L,1} |g, 0, n_L \rangle \\
 = & -\frac{E_{V,k}^* E_\text{L}}{4} \boldsymbol{u}_{k,\pm}^* \cdot \langle g' | \alpha^\textrm{s}_{nL} \boldsymbol{u}_L - i \alpha^\textrm{v}_{nL}  \frac{ \boldsymbol{u}_L \times   \hat{\boldsymbol{J}}}{2J}  |  g \rangle,
\end{align}

Fermi's golden rule gives us the corresponding transition rate
\begin{align}
\Gamma_{g\rightarrow g'} = & \sum_{\boldsymbol{k},\mu} \frac{2 \pi}{\hbar} | V_{g',g}|^2 \delta(E_g + \hbar \omega_L - E_{g'} - \hbar \omega_{\boldsymbol{k}})   \\
 \approx  & \frac{4 \pi |E_L|^2}{16 \hbar \epsilon_0} \sum_k \frac{\hbar c |\boldsymbol{k}|}{2 V \epsilon_0}  \bigg| \boldsymbol{u}_{k,+}^* \cdot \langle g'|    \alpha^\textrm{s}_{nL} \boldsymbol{u}_L - i \alpha^\textrm{v}_{nL}  \frac{ \boldsymbol{u}_L \times   \hat{\boldsymbol{J}}}{2J}   |g \rangle \bigg|^2 \delta(\hbar \omega_L - \hbar \omega_{\boldsymbol{k}}). 
\end{align}

We convert the sum to an integral and integrate over the magnitude of the vacuum mode wavenumbers:
\begin{align}
    \frac{1}{V} \sum_{\boldsymbol{k}} f(k) \rightarrow \frac{1}{(2 \pi)^3} \int d^3k \, f(k) = &{} \frac{1}{(2 \pi)^3} \int d\Omega \, dk \, k^2 f(k) , \\
     \hbar c  \int dk \, k^3 \, \delta(\hbar c (k-k_L)) = & k_L^3 = \frac{\omega_L^3}{c^3}, \\
    I_L = & \frac{c \epsilon_0}{2} |E_L|^2,
\end{align}
which leaves us with the integral over the emission angle,
\begin{equation}
\Gamma_{g\rightarrow g'} = I_L \frac{\omega_L^3}{8 (2 \pi)^2 \hbar c^4 \epsilon_0^2 }\int d\Omega   \bigg| \boldsymbol{u}_{k,\pm}^* \cdot \langle g'|    \alpha^\textrm{s}_{nL} \textbf{u}_L - i \alpha^\textrm{v}_{nL}  \frac{ \boldsymbol{u}_L \times   \hat{\boldsymbol{J}}}{2J}   |g \rangle \bigg|^2 .
\end{equation}
Finally, we make use of the identity
\begin{equation}
    \int d\Omega |\boldsymbol{e}\cdot\boldsymbol{x}|^2 =  \frac{4 \pi}{3}  |\boldsymbol{x}|^2, 
\end{equation}
to get Eq. 1:
\begin{equation}
\Gamma_{g\rightarrow g'} =  I_L \frac{\omega_L^3}{3 \pi \hbar c^4 \epsilon_0^2 }  \bigg| \langle g'|    \alpha^\textrm{s}_{nL} \boldsymbol{u}_L - i \alpha^\textrm{v}_{nL}  \frac{ \boldsymbol{u}_L \times  \hat{\boldsymbol{J}}}{2J}   |g \rangle \bigg|^2.
\end{equation}

\subsection{Trap-mediated Raman adiabatic passage}
The vector component of Eq.~\ref{eq:StarkOperator} takes the form of a fictitious magnetic field 
\begin{equation}
    \boldsymbol{B}^{\text{fict}} = \frac{i \alpha^v_{n J}}{8 \mu_\textrm{B} g_{nJ} J} [  \boldsymbol{\mathcal{E}}^* \times\boldsymbol{\mathcal{E}}],
\end{equation}
where $\mu_\textrm{B}$ is the Bohr magneton and $g_{nJ}$ is the Land\'e factor for the fine-structure level $\ket{n J}$. Consider and electric field with two co-propagating frequency components with relative detuning $\delta$ small enough that the single-frequency expression for the polarizabilities is still approximately valid, and which have orthogonal linear polarizations:
\begin{equation}
     \boldsymbol{\mathcal{E}} =  \boldsymbol{\hat{y}} E_1 e^{i(kx-\omega t)} +  \boldsymbol{\hat{z}} E_2 e^{i(kx-\omega t - \delta t)} .
\end{equation}
In this case the fictitious magnetic field oscillates at the two-photon detuning $\delta$:
\begin{equation}
    \boldsymbol{B}^{\text{fict}} = - \frac{i \alpha^v_{n J}}{8 \mu_\textrm{B} g_{nJ} J}  [ E_1^* E_2 e^{-i \delta t} -  E_1 E_2^* e^{+i \delta t}  ] \boldsymbol{\hat{x}},
\end{equation}
and this term can drive transitions between Zeeman levels in the same way a real oscillating magnetic field can. 

To get from $(3,-3)_{\text{Cs}}$ to $(3,+3)_{\text{Cs}}$ we drive six adiabatic passages. These can be realized by either a downward detuning sweep at fixed magnetic field or an upward magnetic-field sweep at fixed detuning. As the overlap of our beams depends on their detuning due to the acousto-optical modulators used for power stabilization and frequency shifting, we choose a magnetic-field ramp. Lower detunings are somewhat preferable due to the smaller splitting between the transitions, which means that the full transfer is completed in a shorter time given the same magnetic-field ramp speed. We therefore choose a detuning of 41 MHz, which puts the two-photon transitions between 114.25~G and 119.5~G, on the way between 114~G, used for spin purification, and 557~G, used for the final evaporation ramps.

\subsection{Magnetic-field calibration}
\begin{figure}[h!]
  \centering
  \includegraphics[width=0.65\textwidth]{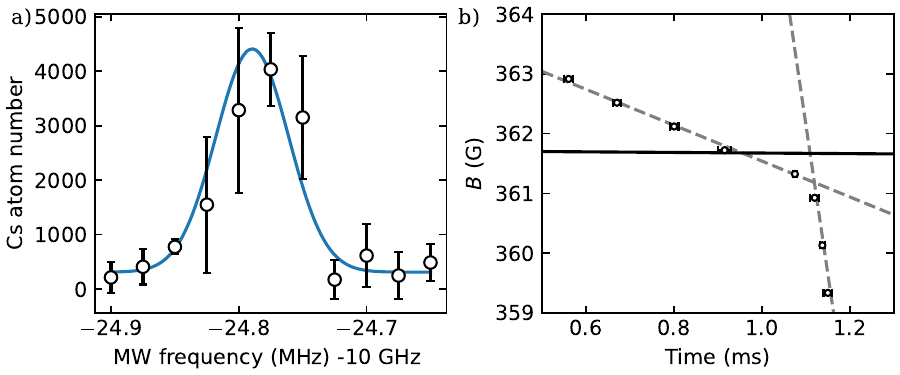}
   \caption{Magnetic-field characterization measurements. (a) A typical magnetic-field calibration measurement, showing the number of detected Cs atoms as a function of the microwave signal frequency. (b) Magnetic field $B$ as a function of time during the magnetoassociation ramp. Zero time corresponds to the moment the coil current is switched to a lower value. The dashed lines correspond to ramp speeds of 3 G/ms and 50 G/ms. The solid horizontal line denotes the position of the Feshbach resonance.}
  \label{fig:associationramp}
\end{figure}

We calibrate our magnetic field by measuring the frequency of the $(f,m_f) = (3,+3)\to (4,+4)$ microwave (MW) transition of Cs. We apply a MW pulse and image the cloud without first flashing the usual repumping light. As a result, we see atoms only when the MW pulse transfers atoms to the $f=4$ manifold. A typical calibration measurement involves measuring atom number versus MW frequency, as shown in Fig.\ \ref{fig:associationramp}(a). We fit a Gaussian function to determine the center of the peak. From this we determine a corresponding magnetic-field value using the Breit-Rabi formula.

Eddy currents in our steel vacuum chamber slow down changes to the magnetic fields set by the coil currents. We therefore measure magnetic-field ramp speeds directly using microwave spectroscopy. For each point along the curve, we fix the microwave frequency to be resonant with the $(3,+3)\to (4,+4)$ transition at a particular magnetic field and measure at which time the adiabatic passage occurs. When performing the magnetoassociation ramp, we quickly reduce the current through one of the coil pairs and rely on the slow decay of the magnetic field to ramp across the Feshbach resonance. Fig.\ \ref{fig:associationramp}(b) shows a measurement of the ramp across the Feshbach resonance. After crossing the resonance we do a magnetic-field jump down to 345~G. The initial rate of change of the magnetic field is about 50~G/cm, which can be considered typical for magnetic-field changes referred to as “jumps” in the main text.

\subsection{Molecular sample characterization}
To count the number of molecules we ramp the magnetic field back up over the Feshbach resonance at 361.7~G to dissociate the molecules. When the molecular state is above the threshold, it quickly decays into a pair of unbound atoms due to the coupling between the molecular state and the continuum of relative motion states of the atom pair. Every molecule is thus converted to an atom pair, and the number of atoms of either species can then be measured using absorption imaging to determine the molecule number before association. For the measurements presented in this work, collisional loss during the dissociation ramp can be neglected due to the low atomic and molecular densities. To measure the temperature of the molecular cloud, we measure the width of its momentum distribution by releasing it from the trap and letting it expand for a variable amount of time before dissociating and imaging the atoms.

\begin{figure}[h!]
  \centering
  \includegraphics[width=0.65\textwidth]{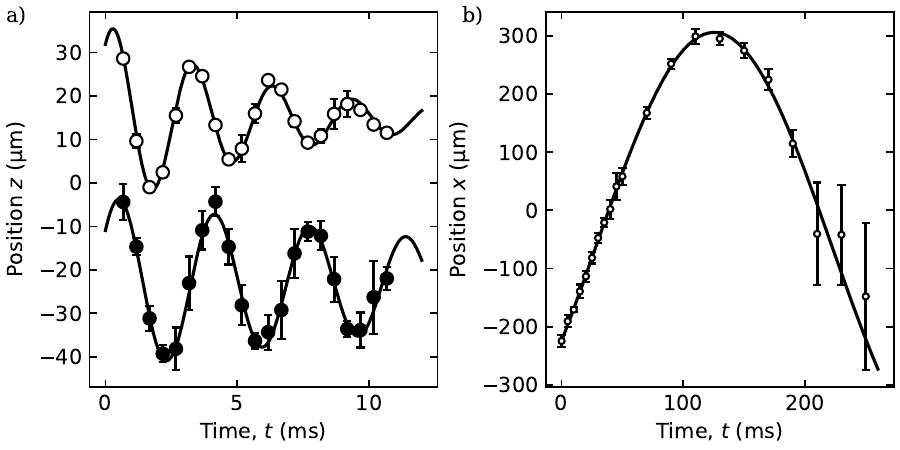}
   \caption{Trap frequency measurements. (a) Vertical displacement in time-of-flight of cold clouds of K (hollow) and KCs Feshbach molecules (filled) in the same trapping conditions. (b) Horizontal oscillations of the KCs Feshbach molecules}
  \label{fig:trapfrequencies}
\end{figure}

The trap potential is a composite of optical dipole trapping perpendicular to the axis of the ODT beam, and magnetic trapping due to magnetic field curvature along the axis of the ODT. We induce oscillations using RODT to measure the radial trap frequency for K atoms and Feshbach molecules (see Fig \ref{fig:trapfrequencies}). Detuning the RF-frequency of the RODT AOM misaligns the beam with respect to the ODT, giving an upward or downward force at the center of the ODT. We find a ratio of $\frac{\omega_{\text{mol}}}{\omega_{\text{K}}} = 0.831(9)$, slightly above the value 0.792 expected from assuming the polarizability of the Feshbach molecule is the sum of polarizabilities of the atoms, and taking into account corrections to the trap frequency due to gravitational sag. Based on the oscillation amplitudes, we estimate the oscillations adds roughly $E_\text{kin} = k_B \times 150\ \text{nK}$ to the molecules and $E_\text{kin} = k_B \times 33\ \text{nK}$ to K atoms, less than one-tenth of the trap depth in both cases.

The switch from 557 G to 345 G and the change in magnetic moment upon molecule creation induces an axial oscillation, which allows us to measure the axial trapping frequency of the molecular cloud. We measure trapping frequencies for the molecules of $f_{\text{rad}} = 195(6)$ Hz and $f_{\text{ax}} = 2.9(3)$ Hz.

 We compute the peak density and phase-space density of the molecules as
\begin{align}
n_0 = & N \bigg( \frac{2 \pi m}{ k_\text{B} T} \bigg)^{3/2} f_{\text{rad}}^2 f_{\text{ax}}, \\
    \text{PSD} = & N \bigg( \frac{h}{k_\text{B} T} \bigg)^3 f_{\text{rad}}^2 f_{\text{ax}}
\end{align}
From the decay curve and the peak density of the cloud we can determine a two-body decay coefficient $\Gamma_2 $. If the main loss process takes the form of inelastic two-body collisions, and the cloud is assumed to stay in thermodynamic equilibrium at fixed temperature, the molecule number decays according to the differential equation
\begin{align}
\dot{N} = & \int d^3 r \, \dot{n}(r) = - \int d^3r \, \Gamma_2 n^2(r) = - \Gamma_2 \alpha N^2,
\end{align}
which has the solution 
\begin{equation}
    N(t) = \frac{N_0}{1+\Gamma_2 \alpha N_0 t},
\end{equation}
where we can substitute $\alpha N_0$ with the mean initial density $\bar{n}_0$ using
\begin{equation}
    \bar{n}_0 = \frac{\int d^3 r \, n^2(r)}{\int d^3 r \, n(r)} = \frac{\alpha N_0^2}{N_0} = \frac{n_0}{2^{3/2}}.
\end{equation}

\subsection{Resonance and bound-state characterization}
\begin{figure}[h!]
  \centering
  \includegraphics[width=\textwidth]{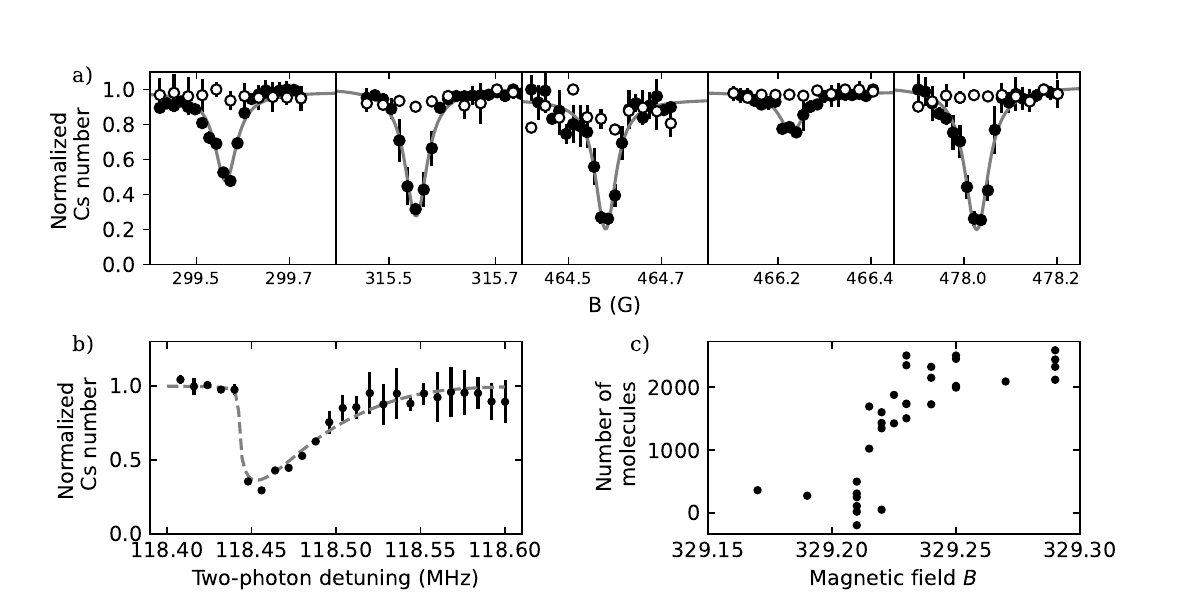}
   \caption{Resonance and bound-state characterization measurements. (a) Loss spectroscopy measurements of the five newly observed resonances. Loss of Cs atoms is observed in the target channel (filled markers), while no loss peak is observed in a different channel with the same Cs state (hollow markers). (b) Bound-state spectroscopy. Remaining Cs fraction as a function of two-photon detuning after 500 ms. The dashed line is a fit of Eq.\ \ref{eq:bindingenergylineshape} (c) Detected molecule number as a function of lowest magnetic field after a slow magnetic-field ramp down followed by a fast ramp up.}
  \label{fig:s3resonancecharacterization}
\end{figure}

Our loss-spectroscopy measurements start from an ultracold mixture of K and Cs atoms. We drive Raman adiabatic passages to prepare the mixture in the channel of interest. Due to eddy currents and magnetization in our steel chamber, the magnetic field needs a time of about 10 to 50 ms to stabilize within 10 mG. Most of the resonances explored in this work are predicted to have widths between 20 and 100 mG. To avoid a significant field shift during the measurement, we first prepare the mixture with K in the target state and Cs in a state with $m_f$ differing by $\pm1$ from the target state. We then change the magnetic field to the target value and hold for 100 ms to let the field stabilize, after which another Raman adiabatic passage transfers Cs to the target state. We hold for an additional 50 to 100 ms, after which the magnetic fields are switched off and we measure the amount of remaining Cs atoms. To verify that the loss features are due to interspecies resonances, we check that they do not appear when K is in a different spin state. A typical loss-spectroscopy measurement is depicted in Fig.\ \ref{fig:s3resonancecharacterization}(a). 

To measure the binding energy of the bound state that corresponds to the resonance at 361.7~G, we perform two-photon spectroscopy. We make use of laser light at 880 nm, the tune-out wavelength of Cs. At this wavelength Cs atoms see no scalar potential from the Raman beams, and the ratio of Raman coupling to spontaneous emission is high, which allows a long probing time. One such binding-energy measurement is shown in Fig.\ \ref{fig:s3resonancecharacterization}(b). The finite temperature of the mixture results in inhomogeneous broadening of the loss spectrum. There is an additional broadening due to the finite lifetime of the molecules. We fit a lineshape of the form $e^{-W(f,f_0)}$ with \cite{PhysRevA.61.012501}
\begin{equation}
W(f,f_0) \propto \int_0^\infty df' \frac{\sqrt{f'}}{(f-f_0-f')^2 + (\Gamma/2)^2}  e^{- b f'} ,
\label{eq:bindingenergylineshape}
\end{equation}
where $f_0 = (E_\textrm{Z}+E_\textrm{b})/h$ is the frequency corresponding to the sum of the Cs Zeeman splitting $E_\textrm{Z}$ and the binding energy $E_\textrm{b}$, $\Gamma$ is the lifetime of the bound state, and $b$ is a constant related to the temperature of the mixture. Close to the pole of the resonance the binding energy takes the universal form $E_\textrm{b} = -\hbar^2/(2 m_r a^2)$, where $m_r$ is the reduced mass of the atom pair and $a\approx -\Delta a_{\text{bg}}/(B-B_\text{res})$ is the scattering length. A fit of these formulas to measured binding energies below 500 kHz yields a resonance position of 361.66(1)~G.

As depicted in Fig.\ \ref{fig:s3resonancecharacterization}(c), we observe loss of our molecular sample when ramping the magnetic field down slowly across 329.21(2)~G, followed by a fast ramp back up. We interpret this as a consecutive adiabatic and diabatic crossing of an avoided crossing between bound states. Molecules created in the s-wave state can be transferred to the d-wave states by means of a sufficiently slow magnetic-field ramp across the avoided crossing. Reversing the slow magnetic-field ramp recovers the s-wave molecules, but a faster ramp leaves molecules in the d-wave state. Our dissociation ramp for imaging s-wave molecules results in a non-adiabatic dissociation of d-wave molecules. As a result, their constituent atoms receive a large amount of kinetic energy, ejecting them from the trap. We have not been able to observe a similar loss feature near 323 G, which we attribute to the combination of a much higher coupling and experimental limitations on maximal magnetic-field ramp speeds. 
 
\subsection{Calculations of bound states and scattering }

We carry out calculations of both bound states and scattering using coupled-channel methods.
Scattering calculations are performed with the MOLSCAT package \cite{molscat:2019, mbf-github:2023}. These calculations produce the scattering matrix $\boldsymbol{S}$, for a single value of the collision energy and magnetic field each time. The complex s-wave scattering length $a$ is obtained from the diagonal element of $\boldsymbol{S}$ in the incoming channel \cite{Hutson:res:2007}. In the present work, s-wave scattering lengths are calculated at $E_\textrm{coll}/k_\textrm{B} = 1$ nK, which is low enough to neglect any dependence on collision energy.

MOLSCAT can converge on Feshbach resonances automatically and characterize them to obtain $B_\textrm{res}$, $\Delta$ and $a_\textrm{bg}$ (and the additional parameters needed in the presence of inelasticity) as described in ref.\ \cite{Frye:resonance:2017}.

Coupled-channel bound-state calculations are performed using the packages BOUND and FIELD \cite{bound+field:2019, mbf-github:2023}, which converge on bound-state energies at fixed field, or bound-state fields at fixed energy, respectively. The methods used are described in ref.\ \cite{Hutson:CPC:1994}. Expectation values are calculated by finite differences, without requiring explicit wave functions \cite{Hutson:expect:88}. In the present paper, this capability is used to calculate overall singlet fractions for bound states.

Zero-energy Feshbach resonances can be fully characterized using MOLSCAT as described above. However, if only the position of the resonance is needed, as when fitting, it is more convenient simply to run FIELD at the threshold energy to locate the magnetic field where the bound state crosses threshold.

In the present work, the coupled equations for both scattering and bound-state calculations are solved using the fixed-step diabatic log-derivative propagator of Manolopoulos \cite{Manolopoulos:1986} from $R_\textrm{min}=5.6\ a_0$ to $R_\textrm{mid}=15\ a_0$, with an interval size of $0.001\ a_0$, and the variable-step Airy propagator of Alexander and Manolopoulos \cite{Alexander:1987} between $R_\textrm{mid}$ and $R_\textrm{max}=3,000\ a_0$.

When considering many different thresholds, it is cumbersome to label them with explicit quantum numbers $(f,m_f)$ or $(m_s,m_i)$, especially as $f$, $m_s$ and $m_i$ are not conserved quantities. In this section we label atomic states instead with Roman letters a to h, in increasing order of energy, and label thresholds with pairs of letters with the one for $^{39}$K first.

The only conserved quantities in a magnetic field are $M_\textrm{tot} = m_{f,\textrm{K}} + m_{f,\textrm{Cs}} + M_L = M_F + M_L$ and parity $(-1)^L$. We take advantage of this to perform calculations for each $M_\textrm{tot}$ and parity separately. In each calculation, we include all basis functions of the required $M_\textrm{tot}$ and parity for $s_\textrm{K}=s_\textrm{Cs}=\frac{1}{2}$, $i_{^{39}\textrm{K}}=\frac{3}{2}$ and $i_\textrm{Cs}=\frac{7}{2}$, subject to the limitation $L\le L_\textrm{max}$. In most of the calculations in the present work, $L_\textrm{max}=0$, except that we use $L_\textrm{max}=2$ for the bound states in Fig.\ 4. When $L_\textrm{max}=0$, $M_F=M_\textrm{tot}$, so $M_F$ is conserved.

For the calculations in Fig.\ 4, the singlet and triplet potentials are supplemented with the spin-spin coupling operator $\hat{V}^\textrm{d}(R)$ of ref.\ \cite{Patel:2014}, which provides weak coupling between $L=0$ and $L=2$ and is responsible for the corresponding avoided crossings in Fig.\ 4. It has no effect in calculations that include $L=0$ only.

\begin{figure}[b]
\begin{center}
\includegraphics[width=0.62\textwidth]{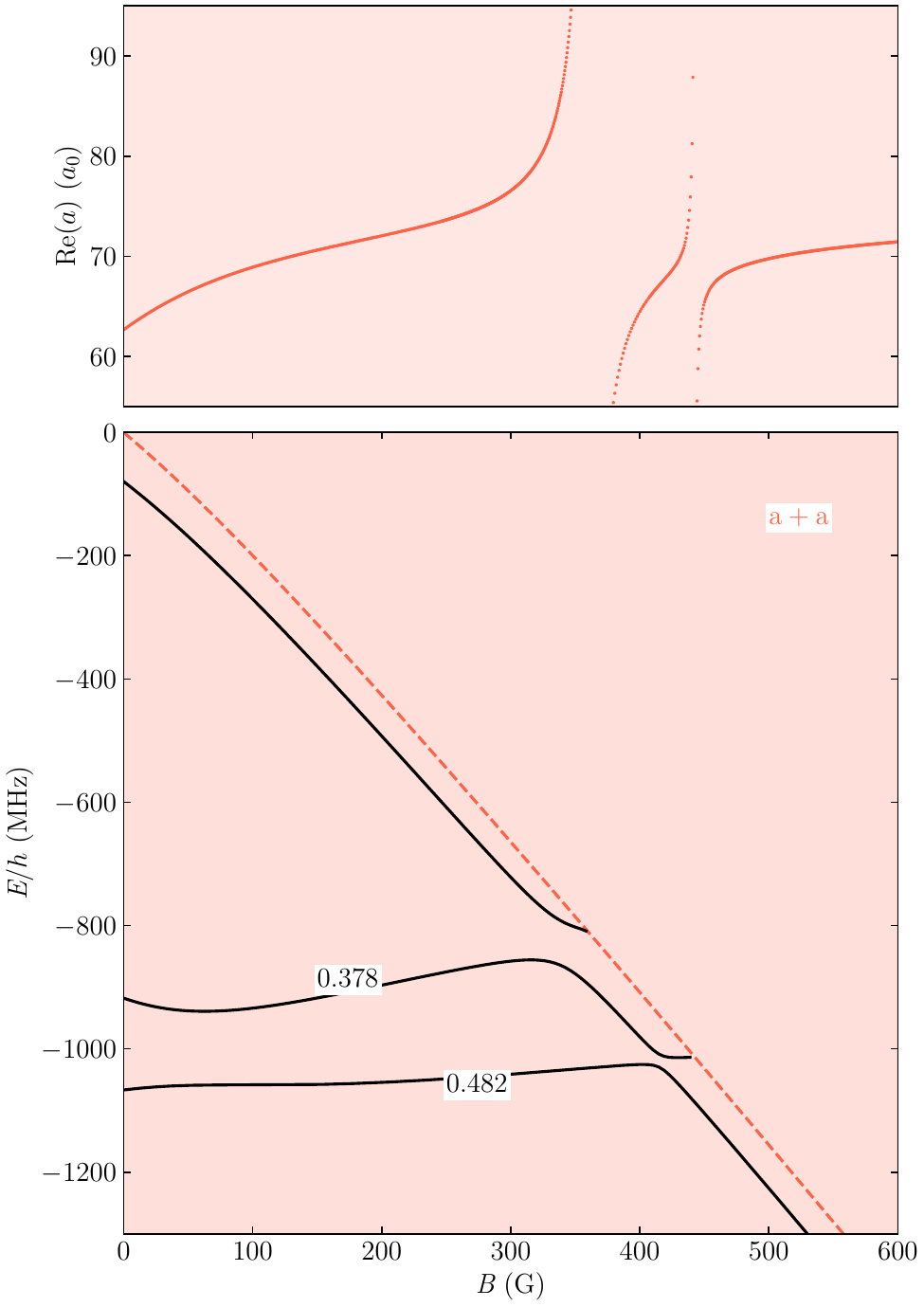}
\caption{Bound states of $^{39}$KCs with $M_\textrm{tot}=4$ as a function of magnetic field $B$, showing their crossings with the lowest threshold (a+a, dashed line). Each state is labeled with its singlet fraction in a region far from avoided crossings. Upper panel: scattering length at the lowest threshold. The results are from coupled-channel calculations on the potentials of ref.\ \cite{GrobnerKCsFeshbach} including basis functions with $L=0$ only.}
\label{fig:MF4}
\end{center}
\end{figure}

\begin{figure}[t]
\begin{center}
\includegraphics[width=0.70\textwidth]{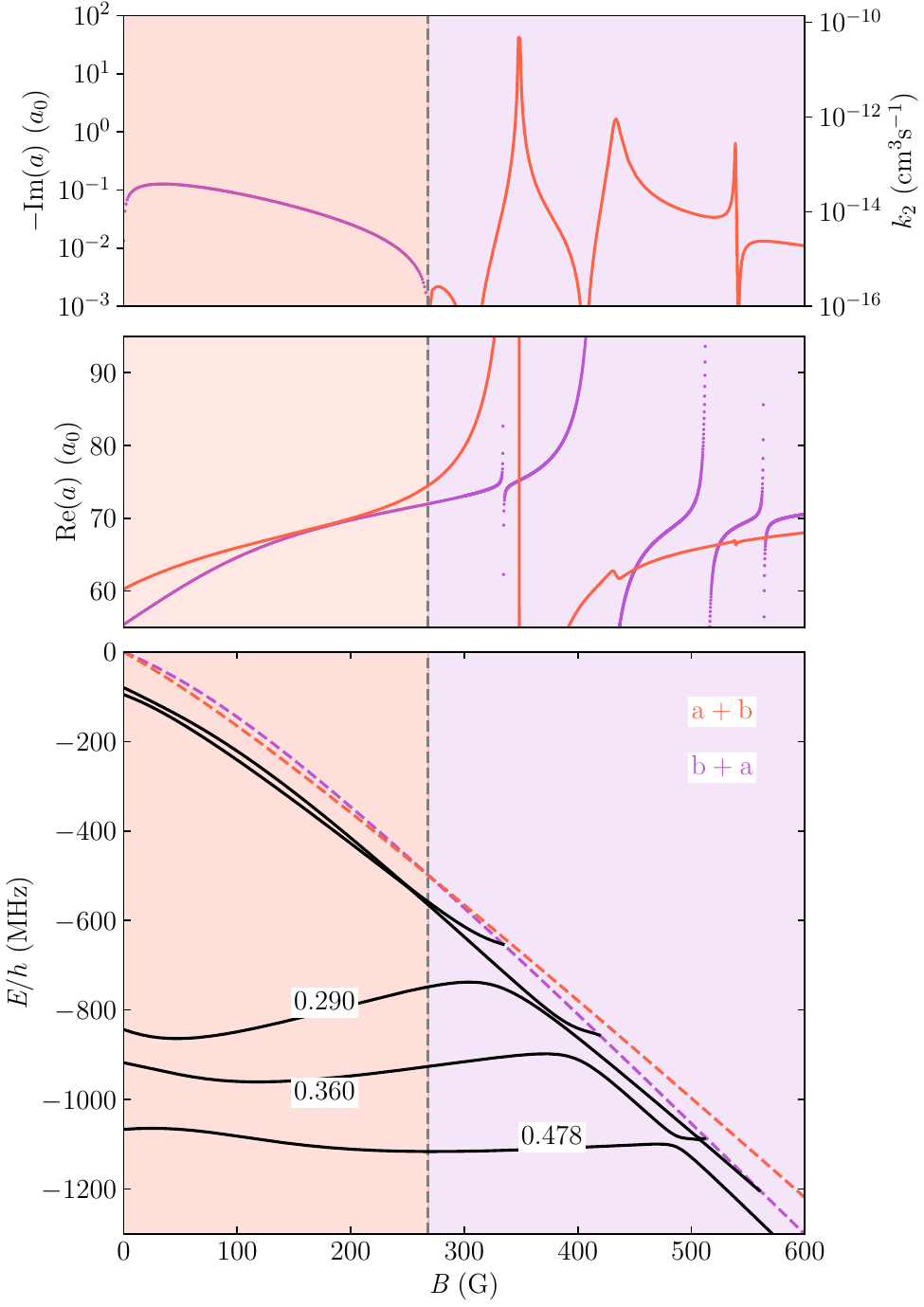}
\caption{Bottom panel: Bound states of $^{39}$KCs with $M_F=3$ (black lines) as a function of magnetic field $B$, with the thresholds for $M_F=3$ (dashed lines) colour-coded as shown in the legend. Coloured shading indicates the region in which each threshold is the lowest for this $M_F$. Each state is labeled with its singlet fraction in a region far from avoided crossings. Middle panel: real part of scattering length at each threshold. Top panel: imaginary part of scattering length at each threshold for which inelastic collisions are possible, with right-hand axis showing the 2-body loss rate at limitingly low collision energy. The results are from coupled-channel calculations on the potentials of ref.\ \cite{GrobnerKCsFeshbach} including basis functions with $L=0$ only.}
\label{fig:MF3}
\end{center}
\end{figure}

\begin{figure}[t]
\begin{center}
\includegraphics[width=0.70\textwidth]{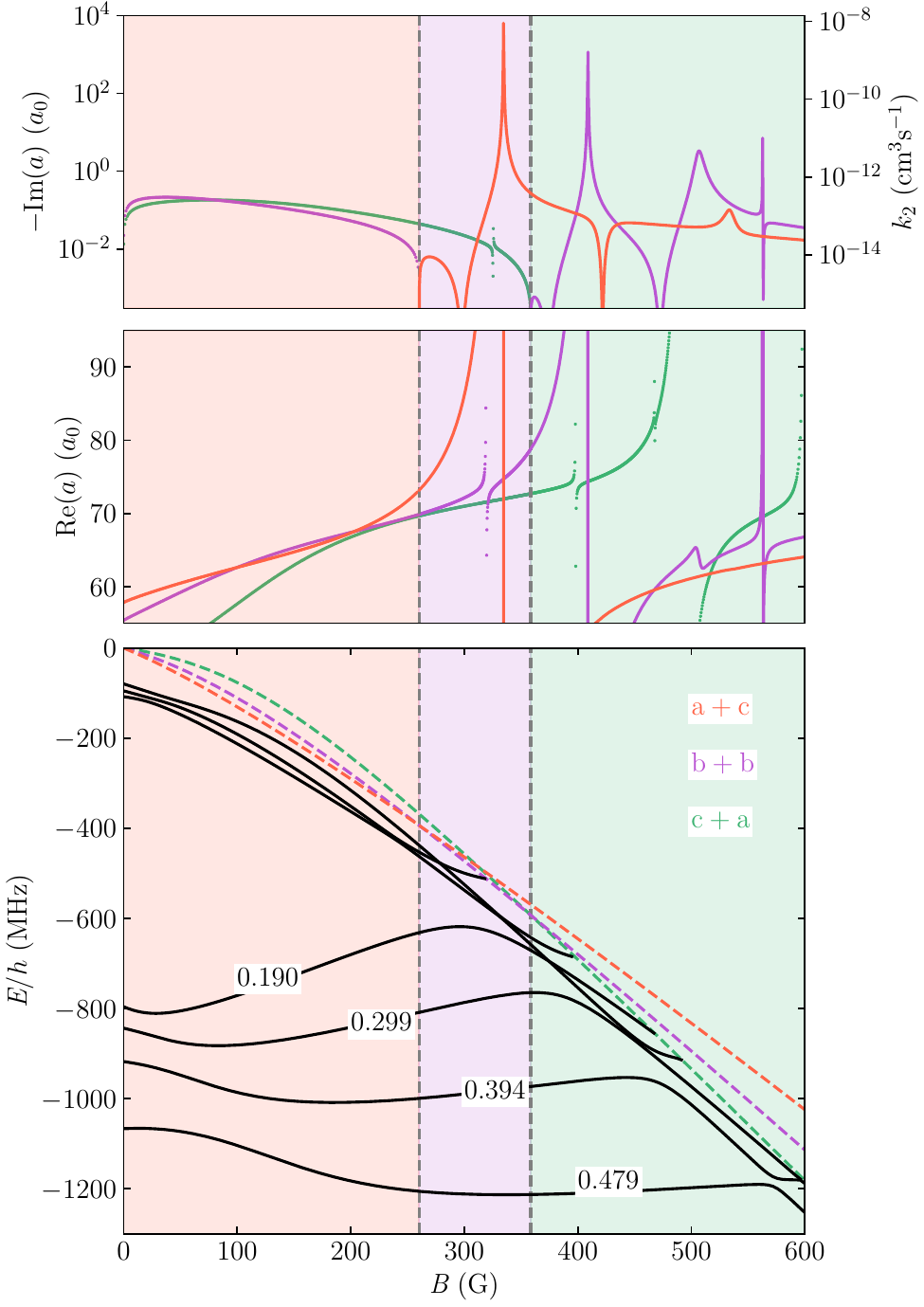}
\caption{Bottom panel: Bound states of $^{39}$KCs with $M_F=2$ (black lines) as a function of magnetic field $B$, with the thresholds for $M_F=2$ (dashed lines) colour-coded as shown in the legend. Coloured shading indicates the region in which each threshold is the lowest for this $M_F$. Each state is labeled with its singlet fraction in a region far from avoided crossings. Middle panel: real part of scattering length at each threshold. Top panel: imaginary part of scattering length at each threshold for which inelastic collisions are possible, with right-hand axis showing the 2-body loss rate at limitingly low collision energy. The results are from coupled-channel calculations on the potentials of ref.\ \cite{GrobnerKCsFeshbach} including basis functions with $L=0$ only.}
\label{fig:MF2}
\end{center}
\end{figure}

\begin{figure}[t]
\begin{center}
\includegraphics[width=0.70\textwidth]{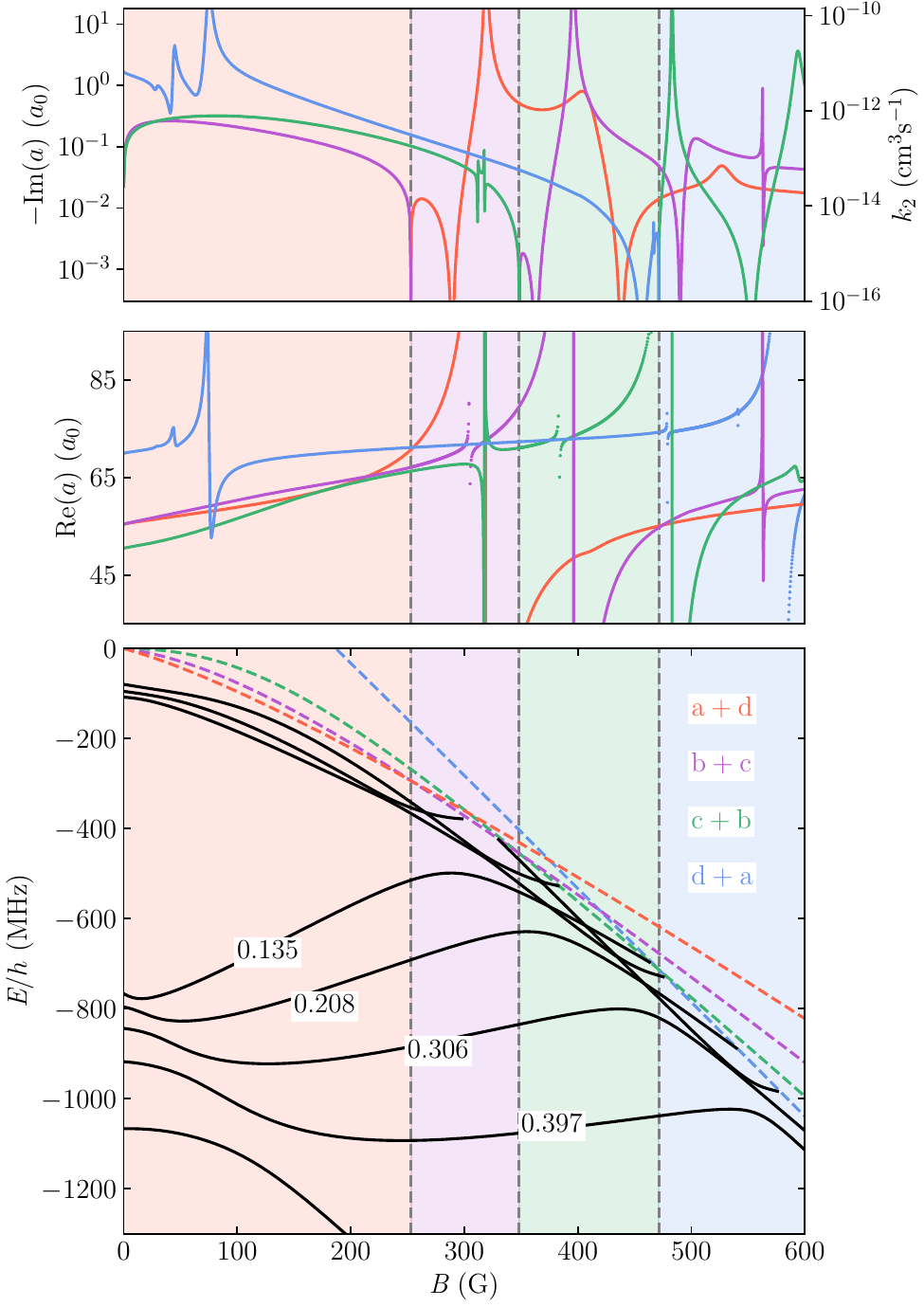}
\caption{Bottom panel: Bound states of $^{39}$KCs with $M_F=1$ (black lines) as a function of magnetic field $B$, with the thresholds for $M_F=1$ (dashed lines) colour-coded as shown in the legend. Coloured shading indicates the region in which each threshold is the lowest for this $M_F$. Each state is labeled with its singlet fraction in a region far from avoided crossings. Middle panel: real part of scattering length at each threshold. Top panel: imaginary part of scattering length at each threshold for which inelastic collisions are possible, with right-hand axis showing the 2-body loss rate at limitingly low collision energy. The results are from coupled-channel calculations on the potentials of ref.\ \cite{GrobnerKCsFeshbach} including basis functions with $L=0$ only.}
\label{fig:MF1}
\end{center}
\end{figure}

\begin{figure}[t]
\begin{center}
\includegraphics[width=0.70\textwidth]{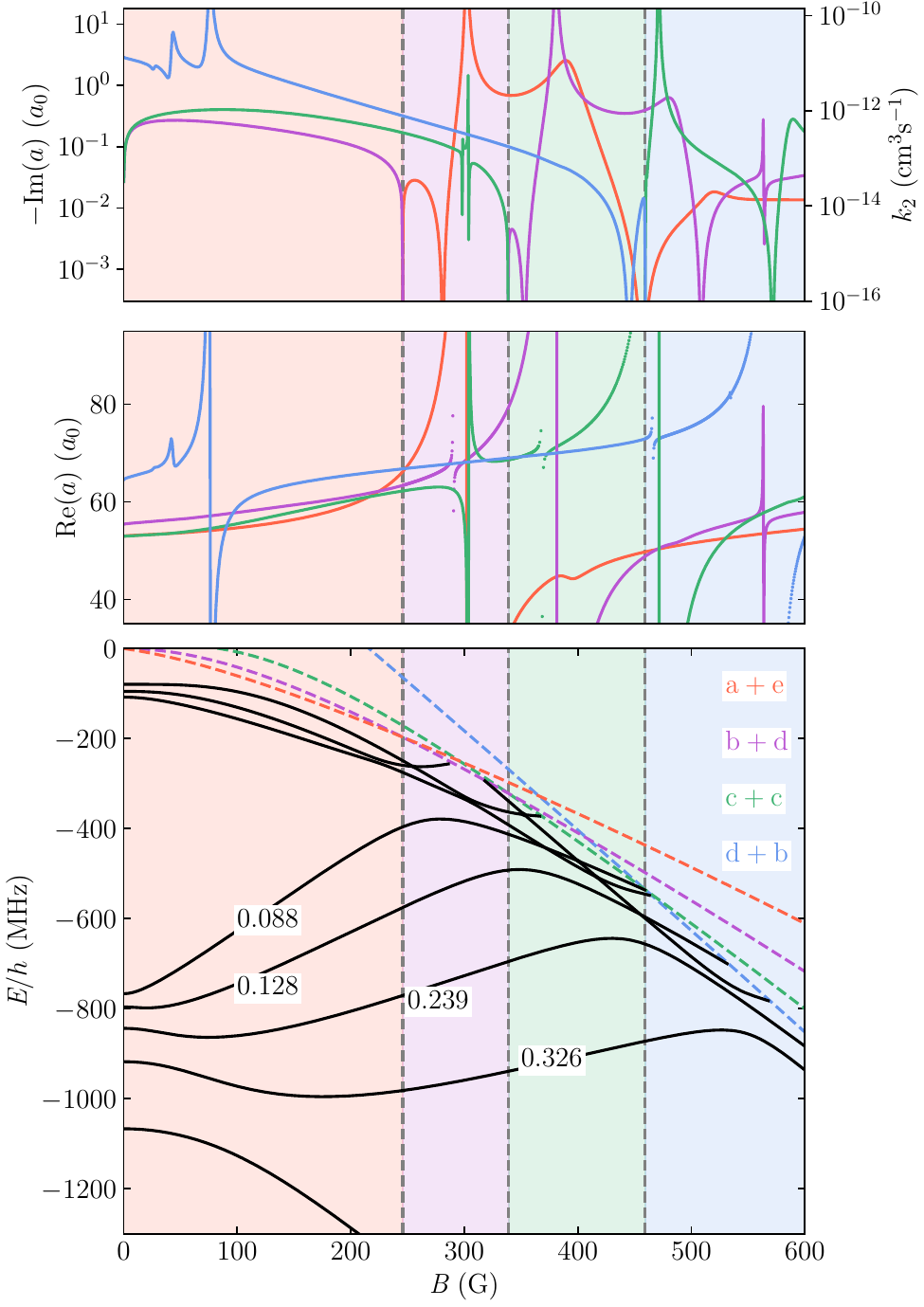}
\caption{Bottom panel: Bound states of $^{39}$KCs with $M_F=0$ (black lines) as a function of magnetic field $B$, with the thresholds for $M_F=0$ (dashed lines) colour-coded as shown in the legend. Coloured shading indicates the region in which each threshold is the lowest for this $M_F$. Each state is labeled with its singlet fraction in a region far from avoided crossings. Middle panel: real part of scattering length at each threshold. Top panel: imaginary part of scattering length at each threshold for which inelastic collisions are possible, with right-hand axis showing the 2-body loss rate at limitingly low collision energy. The results are from coupled-channel calculations on the potentials of ref.\ \cite{GrobnerKCsFeshbach} including basis functions with $L=0$ only.}
\label{fig:MF0}
\end{center}
\end{figure}

Figures \ref{fig:MF4} to \ref{fig:MF0} show the bound-state and threshold energies for each $M_F$ from 4 (which includes the lowest threshold, a+a) to 0. These calculations are on the potentials of ref.\ \cite{GrobnerKCsFeshbach} (before the present fit), but are sufficient for understanding the nature of the levels that cause the resonances observed here. They use basis sets with $L_\textrm{max}=0$, which (for $M_F<4$) is required to make the states below the lowest threshold bound rather than quasibound. In every case there is a bound state that runs almost parallel to each threshold, about 80 MHz below it. These are the least-bound states in each channel, with quantum number $n=-1$ with respect to threshold. They are crossed by sets of states that have very different gradients; these are states supported principally by higher thresholds, with $n=-2$. They have strong avoided crossings with the $n=-1$ states, which are generally not complete where the mixed state crosses the lowest threshold. Nevertheless, the fields where they cross threshold are principally determined by the energies of the states with $n=-2$, which are each labeled with their singlet fraction $F_\textrm{s}$ in a region far from avoided crossings. The corresponding triplet fraction is $F_\textrm{t}=1-F_\textrm{s}$, and the sensitivity of the resonance positions to the singlet and triplet potentials is governed largely by $F_\textrm{s}$ and $F_\textrm{t}$.

The upper two panels of Fig.\ \ref{fig:MF4} to \ref{fig:MF0} show the real and imaginary parts of the field-dependent scattering length $a(B)$. Resonances at the lowest threshold (for each $M_F$) occur as poles in Re($a$), with Im($a$) zero, while resonances at higher thresholds appear as finite oscillations in both Re($a$) and Im($a$) \cite{Hutson:res:2007}. The panels for Im($a$) show the corresponding low-energy 2-body loss rate on the right-hand axis. It is clear that, at thresholds that are not the lowest-energy one for their $M_F$, even the background 2-body loss is fast in most cases, so that observing resonances by loss spectroscopy at these thresholds would be challenging. 

There are a few resonances that occur where states with $n=-1$ cross threshold without significant perturbation from states with $n=-2$. These occur, for example, near 564 G at the b+a threshold and near 448 G at the c+a threshold. Measuring their positions is valuable because the difference in binding energies between states with different $n$ is very sensitive to the dispersion coefficient $C_6$ \cite{Brookes:2022}.

After fitting, we characterized all the resonances that have widths $\Delta>10^{-3}$~G at fields below 600~G at the lowest threshold for each $M_F$ with coupled-channel calculations on the fitted potential. They are summarized in Table \ref{tab:B2024-res-predictions}.

\clearpage
\begin{table}[tb]
\centering
      \caption{Resonances with widths $\Delta>10^{-3}$~G at fields below 600~G at the lowest threshold for each $M_F$, characterized with coupled-channel calculations including basis functions with $L=0$ only. Experimental values for the resonance positions are given for the resonances measured in ref.~\cite{GrobnerKCsFeshbach} and in this work.} 
\begin{tabular}{cccccc}
\hline\hline
 $\mathrm{Threshold}$ & $M_F$ & $B_\mathrm{res}$ (G) & $\Delta$ (G) & $a_{\mathrm{bg}}$ ($a_0$) & $B_\text{exp}^\text{loss}$ (G) \\
\hline
a+a & 4 & 361.66 & 5.23 & 74.65  & 361.66(1) \\
a+a & 4 & 443.45 & 0.41 & 70.08  & 442.6(3) \cite{GrobnerKCsFeshbach} \\
b+a & 3 & 332.23 & 0.03 & 77.60  & \\
b+a & 3 & 420.24 & 4.80 & 74.74  & 419.3(3) \cite{GrobnerKCsFeshbach} \\
b+a & 3 & 513.77 & 0.58 & 71.29  & 513.1(3) \cite{GrobnerKCsFeshbach}\\
b+a & 3 & 553.58 & 0.09 & 71.47  & \\
b+b & 2 & 315.77 & 0.07 & 76.60  & 315.551(2) \\
c+a & 2 & 396.36&  0.03 & 77.05  & \\
c+a & 2 & 465.61 & 7$\times 10^{-3}$ & 86.46  & 466.230(5) \\
c+a & 2 & 492.24 & 4.23 & 74.89  & 491.5(3) \cite{GrobnerKCsFeshbach}\\
c+a & 2 & 597.78 & 0.46 & 72.85  & 599.3(3) \cite{GrobnerKCsFeshbach}\\
c+a & 2 & 609.04 & 0.10 & 69.63  & \\
b+c & 1 & 299.87 & 0.08 & 74.12  & 299.562(4) \\
c+b & 1 & 380.67 & 0.07 & 75.89   & \\
d+a & 1 & 478.07 & 0.02 & 77.38  & 478.028(4) \\
d+a & 1 & 538.91 & $2\times 10^{-3}$ & 81.87   & \\
d+a & 1 & 578.88 & 3.66 & 75.23   & \\
b+d & 0 & 284.43 & 0.07 & 69.90  & \\
c+c & 0 & 364.58 & 0.09 & 73.41  & \\
d+b & 0 & 464.63 & 0.04 & 76.12   & 464.579(3) \\
d+b & 0 & 534.03 & $2\times 10^{-3}$ & 87.24   & \\
d+b & 0 & 571.11 & 8.17 & 71.90   & \\
\hline\hline
\end{tabular}
\label{tab:B2024-res-predictions}
\end{table}

\end{document}